\documentclass[11pt]{article}
\usepackage[margin=1in]{geometry} 
\usepackage{amsmath,amsthm,amssymb,amsfonts, titlesec, physics, enumitem, accents, setspace, fancyhdr, geometry, pdflscape}
\usepackage{graphicx}
\usepackage{float}
\usepackage[font=footnotesize,labelfont=bf]{caption}
\usepackage{array}
\usepackage[title]{appendix}
\usepackage{afterpage}
\usepackage{listings}
\usepackage{booktabs}
\usepackage[hidelinks]{hyperref}
\usepackage{natbib}
\usepackage{xcolor}
\usepackage{bbm}

\titleformat{\section}{\scshape\bfseries}{\thesection.}{1em}{}
\titleformat{\subsection}{\normalfont\itshape}{\thesubsection.}{1em}{}

\newcommand{\midd}{\,|\,}
\newcommand{\I}{\mathbbm{1}}

\allowdisplaybreaks

\begin{document}
\begin{center}
    \textbf{\Large{Investigating symptom duration using current status data: \\a case study of post-acute COVID-19 syndrome 
}} \vspace{0.5in}

Charles J. Wolock\textsuperscript{*1}, Susan Jacob\textsuperscript{*2}, Julia C. Bennett\textsuperscript{2}, Anna Elias-Warren\textsuperscript{2}, Jessica O'Hanlon\textsuperscript{2}, Avi Kenny\textsuperscript{3,4}, Nicholas P. Jewell\textsuperscript{5,6}, Andrea Rotnitzky\textsuperscript{7}, Stephen R. Cole\textsuperscript{8}, Ana A. Weil\textsuperscript{9,10}, Helen Y. Chu\textsuperscript{2} \& Marco Carone\textsuperscript{7}  \vspace{0.4in}

\begin{footnotesize}
\textsuperscript{1}Department of Biostatistics, Epidemiology and Informatics, University of Pennsylvania\\
\textsuperscript{2}Department of Epidemiology, University of Washington\\
\textsuperscript{3}Department of Biostatistics and Bioinformatics, Duke University\\
\textsuperscript{4}Global Health Institute, Duke University\\
\textsuperscript{5}Department of Medical Statistics, London School of Hygiene and Tropical Medicine\\
\textsuperscript{6}Division of Biostatistics, University of California, Berkeley\\
\textsuperscript{7}Department of Biostatistics, University of Washington\\
\textsuperscript{8}Department of Epidemiology, University of North Carolina at Chapel Hill\\
\textsuperscript{9}Department of Medicine, University of Washington\\
\textsuperscript{10}Department of Global Health, University of Washington \\
\end{footnotesize}
\vspace{0.3in}

\textbf{Abstract}
\begin{quote}
    For infectious diseases, characterizing symptom duration is of clinical and public health importance. Symptom duration may be assessed by surveying infected individuals and querying symptom status at the time of survey response. For example, in a SARS-CoV-2 testing program at the University of Washington, participants were surveyed at least $28$ days after testing positive and asked to report current symptom status. This study design yielded current status data: outcome measurements for each respondent consisted only of the time of survey response and a binary indicator of whether symptoms had resolved by that time. Such study design benefits from limited risk of recall bias, but analyzing the resulting data necessitates tailored statistical tools. Here, we review methods for current status data and describe a novel application of modern nonparametric techniques to this setting. The proposed approach is valid under weaker assumptions compared to existing methods, allows use of flexible machine learning tools, and handles potential survey nonresponse. From the university study, under an assumption that the survey response time is conditionally independent of symptom resolution time within strata of measured covariates, we estimate that $19$\% of participants experienced ongoing symptoms $30$ days after testing positive, decreasing to $7$\% at $90$ days. We assess the sensitivity of these results to deviations from conditional independence, finding the estimates to be more sensitive to assumption violations at 30 days compared to 90 days. Female sex, fatigue during acute infection, and higher viral load were associated with slower symptom resolution.  \vspace{0.2in}
    
\end{quote}
\end{center}

\begin{NoHyper}
    \def\thefootnote{*}\footnotetext{The first two authors contributed equally to this work.}
\end{NoHyper}
\newpage
\doublespacing

\section*{Introduction}
The persistence of COVID-19 symptoms is a feature of the natural history of SARS-CoV-2 infection. Early in the COVID-19 pandemic, it became clear that many infected individuals suffered symptoms continuing beyond the acute phase of the disease \citep{goertz2020persistent, liang2020three, van2021comprehensive}. Post-acute sequelae are not unique to SARS-CoV-2; many viruses, including the Ebola, Epstein-Barr, chikungunya, dengue, polio, and SARS viruses, among others, have been linked to post-acute infection syndromes \citep{choutka2022unexplained}. There is now a substantial body of literature on the post-acute sequelae of COVID-19 (PASC), colloquially known as long COVID, but there is no universally agreed upon case definition for PASC. Recently, the National Academies proposed defining PASC as an infection-associated chronic condition persisting for at least three months after acute SARS-CoV-2 infection \citep{nasem}. The World Health Organization instead suggests that a patient can be considered to experience PASC if, three months after infection, they have experienced symptoms lasting at least two months that are not explainable by an alternative diagnosis \citep{soriano2022clinical}.  
	
Since there is no consensus on the duration of symptoms warranting a PASC diagnosis, it may be informative to study the distribution of time from SARS-CoV-2 infection --- or, more feasibly, a suitable proxy such as onset of symptoms or confirmation of infection by testing --- until symptom resolution. In this article, we are motivated by the problem of studying the duration of COVID-19 symptoms using data collected as part of the Husky Coronavirus Testing (HCT) initiative. From September $2020$ to June $2023$, the HCT study enrolled 40,000 participants at the University of Washington (UW) in Washington state \citep{weil2021sars}. Beginning in December $2021$, roughly at the beginning of the initial wave of Omicron variant infections, participants who reported a SARS-CoV-2 infection confirmed by either polymerase chain reaction (PCR) test or at-home rapid antigen test were invited to complete a follow-up survey regarding the persistence of symptoms in the months following infection. The aim of this survey was to estimate how common persistent post-acute COVID-19 symptoms are, and to identify risk factors for persistence in a population of university affiliates. Notably, this population is predominantly young and highly vaccinated, with relatively few preexisting comorbidities. In the survey, participants were asked whether, at the time of survey completion, they continued to experience symptoms associated with their recent infection.

The exact time from infection until symptom resolution was not observed for any participant in the HCT long COVID study. Instead, for each survey respondent, in addition to baseline covariate information collected at or before the test date, the available data consist of the time elapsed from positive SARS-CoV-2 test until survey response --- referred to as the survey response time --- and a binary variable indicating whether the respondent’s symptoms had resolved as of the survey response time --- referred to as the event indicator. Each participant was observed only at a single survey response time. Response times in this study ranged from $28$ days (the earliest time after a positive test at which surveys were sent) to $257$ days. Because by design investigators recorded only the current status of a participant’s symptom persistence at their response time, this sampling scheme, which represents an extreme form of interval censoring, is referred to as current status sampling.

Current status data arise and have been analyzed in a number of epidemiological settings, including partner studies of HIV transmission \citep{jewell1990partner}, cross-sectional studies for estimating the age of onset of a non-fatal disease \citep{keiding1991age, keiding1996estimation}, and demographic studies of age at weaning \citep{grummer1993regression, lutter2011increases, rinaldi2019agreement, khan2020breastfeeding, harvey2022maternal}. Designs analogous to the HCT long COVID study, in which participants are surveyed following a health-related event and queried regarding the presence of sequelae, have been used, for example, to study outcomes following radiation therapy \citep{md2017radiation}, joint surgery \citep{pinsker2016ankle, lim2022knee}, diagnosis with light chain amyloidosis \citep{d2023lightchain}, onset of schizophrenia \citep{fan2021schizophrenia}, and infection with Ebola virus \citep{mattia2016ebola}. In fact, previous studies of persistent COVID-19 symptoms have also used a similar design \citep{ganesh2021promis, robineau2022long, imoto2022cross, phen2024prevalence}. Such designs may be appealing because they are simple to implement, economical, and most importantly, less susceptible to recall bias than retrospective designs in which participants are asked to recall symptom status at a specific date in the past \citep{landry2023postacute}. Nevertheless, the presence of current status data may be easily missed; many of the aforementioned studies treated occurrence of the outcome as a binary variable, ignoring variation in response times. Recognizing the presence of current status data is vital, since their analysis requires specialized statistical techniques \citep{jewell2003current}.

Traditional analysis techniques for current status data require that survey response times be uninformative of the symptom resolution process. In this paper, we show that recent methodological advances we have made in causal inference under shape constraints can be adapted to make statistical inference on the time-to-symptom-resolution distribution allowing some response time informativeness. The proposed approach allows us to use flexible machine learning tools and thereby minimize the need for restrictive modeling assumptions. Furthermore, our method accounts for potential nonresponse, which is ubiquitous in survey studies such as the HCT long COVID study, where over half of survey recipients had not responded when the study concluded. To our knowledge, the link between the methodology we build upon here, termed causal isotonic regression (CIR), and the analysis of current status data has not been made before; we aim to highlight the utility of this approach. Using our method, we provide substantive results regarding persistent COVID-19 symptoms in the HCT population during the early Omicron period of the SARS-CoV-2 pandemic. To accompany these results, we devise a sensitivity analysis approach investigating the degree to which the estimated time-to-symptom-resolution distribution changes under deviations from our key assumption. We also investigate baseline patient characteristics associated with time to resolution, that is, risk factors for delayed symptom resolution, using existing regression methods for current status data.

\section*{Methods}
\subsection*{Current status data structure}
We begin by describing in general the current status data structure we are considering, using the HCT study for concreteness. For each recipient of the long COVID survey, a collection $W$ of baseline covariates, observed at or before the time of a positive SARS-CoV-2 test, were recorded. Estimating the distribution of symptom duration $T$, the time elapsed between infection and symptom resolution, is challenging because ($1$) $T$ is never observed directly and ($2$) the study is terminated before all participants have responded to the survey. 

In an ideal study with unlimited follow-up, we would observe for each participant the survey response time $Y_*$ --- the time elapsed between positive SARS-CoV-2 test and survey response --- and the event indicator $\Delta_*$ --- a participant's symptom status at the time of survey response --- with $\Delta_* = 1$ representing symptom resolution by time $Y_*$ and $\Delta_* = 0$ otherwise. We denote by $c_0$ the investigator-determined time (since the positive test) at which follow-up ends for a participant, and consider as a nonrespondent any participant without a response by time $c_0$. The value of $c_0$ is the same for all participants in the study. We set $Y = \min(Y_*, c_0)$ as the smaller of the survey response time and $c_0$, and denote by $\Delta = \I(Y_* < c_0)\Delta_*$  the observed response indicator, which equals $1$ if a participant responded and reported resolution and $0$ otherwise. The available data are considered to be $n$ independent and identically distributed observations $(W_1, Y_1, \Delta_1), (W_2, Y_2, \Delta_2), \ldots, (W_n, Y_n, \Delta_n)$. 

In statistical terms, the primary goal of our analysis consists of using the available data to estimate the survival function of $T$. It may also be of interest to estimate the association between baseline covariates $W$ and symptom duration $T$. Below, we describe how these two goals may be achieved.

\subsection*{Statistical procedure}
For a given time $t$, we denote by $\psi(t)=P(T>t)$ the survival function of $T$ evaluated at $t$. This quantity represents the proportion of individuals testing positive for SARS-CoV-2 who experience symptom persistence beyond time $t$ days. Our aim is to estimate $\psi(t)$ over an interval $[t_0,t_1]$ strictly contained in $(b_0, c_0)$, where $b_0 = 28$ days is the smallest observable response time. Estimating this function using current status data requires tailored statistical tools, since failure to account for the sampling scheme can lead to misleading conclusions. 

Traditionally, under current status sampling, the survival function has been estimated based on the nonparametric maximum likelihood estimator (NPMLE), which can be computed using the pool adjacent violators algorithm  used for isotonic (i.e., monotone) regression \citep{ayer1955empirical}. The key idea behind this approach is that estimation of the survival function can be reframed as a regression problem under a monotonicity constraint: the probability of symptom resolution by time $t$ is simply the proportion of individuals who have resolved symptoms among those with survey response time $t$, and this probability is necessarily nondecreasing as a function of survey response time. This estimator has the advantage of being distribution-free, in the sense that its validity does not rely on any assumption about the underlying shape of the true survival function, and of enabling the construction of valid confidence intervals derived from its large-sample properties \citep{groeneboom1987asymptotics}. Nevertheless, an important limitation of the NPMLE approach is that it requires the survey response process to be entirely uninformative, in the sense that the response time itself does not carry information about the event time. This condition may fail to hold in practical settings, as there may exist baseline covariates --- analogous to confounders in a causal inference setting --- that inform the relationship between the event indicator and the survey response time. For example, participants with substantial comorbidities may be more likely to respond promptly to the survey and to have greater symptom persistence. We argue here that the framework of CIR introduced in \citet{westling2020causal}, which was developed to estimate a monotone causal dose-response curve, can be adapted to estimate the survival function with current status data under weaker conditions. Specifically, as opposed to the traditional NPMLE approach, this method only requires independence between the survey response and event times \textit{within strata} defined by the recorded baseline covariates. Such independence condition is typically much more realistic, and thereby makes the analysis less likely to yield misleading conclusions due to assumption violations.

To describe how CIR can be used to analyze current status data, we introduce the potential outcome notation $\Delta_*(t)$ to denote the value that $\Delta_*$ would take were $Y_*$ set to value $t$, e.g., in a hypothetical setting where the investigator controlled the survey response time. We note that, among survey respondents, the observed event indicator $\Delta$ corresponds to $\Delta_*(Y)$, since by definition $Y < c_0$ for respondents. The survival function evaluated at time $t$ can then be written as $\psi(t)=1-E\{\Delta_*(t)\}$. This function is necessarily monotone since the probability of symptom persistence at survey response time $Y_* = t$ cannot increase in $t$. If $\Delta_*(t)$ and $Y_*$ are conditionally independent given $W$, then the target parameter evaluated at any $t$ in $[t_0,t_1]$ can be written as
\begin{alignat*}{3}
    &\psi(t)\ &&=\ 1-E\{\Delta_*(t)\} &&\text{(by definition)} \\
 &&&=\ 1-E[E\{\Delta_*(t) \midd W\}]&&\text{(by law of total expectation)}\\
&&&=\ 1-E[E\{\Delta_*(t) \midd W,Y_*=t\}]\hspace{.5in}&&\text{(by conditional independence)}\\
&&&=\ 1-E[E\{\Delta_*(t) \midd W,Y=t\}]\hspace{.5in}&&\text{(since $Y_* = t$ if and only if $Y = t$ for $t < c_0$)}\\
&&&=\ 1-E\left\{E\left(\Delta \midd  W,Y = t\right)\right\}&&\text{(since $\Delta = \Delta_*(Y)$)}\ .
\end{alignat*}
Therefore, under this conditional independence assumption, the task of estimating $\psi(t)$ is equivalent to estimation of what may be viewed as a monotone causal dose-response curve, where the exposure and potential outcomes are assumed to be independent conditional on measured covariates. 

Computation of the NPMLE involves performing an isotonic regression of $\Delta_1,\Delta_2, \ldots, \Delta_n$ on $Y_1, Y_2, \ldots, Y_n$ over the domain $[t_0,t_1]$. Computation of the CIR estimator is more complex since it includes a step to disentangle the dependence between event and survey response times. This step involves estimation of two functions depending on baseline covariates and survey response times: ($1$) $\mu(y,w)=E(\Delta \midd Y=y,W=w)$, the conditional mean of $\Delta$ given $Y$ and $W$; and ($2$) $g(y,w)=\pi(y \midd w)/E\{\pi(y\midd W)\}$, a standardization of the conditional density $\pi$ of $Y$ given $W$. These functions play a role akin to those of the outcome regression and (standardized) propensity score in causal inference for continuous exposures. To implement the CIR procedure, estimates $\mu_n$ and $g_n$ of $\mu$ and $g$, respectively, are first constructed; this task can be accomplished with any off-the-shelf machine learning algorithm. Pseudo-outcomes $\Gamma_n (\Delta_1,Y_1,W_1 ),\Gamma_n (\Delta_2,Y_2,W_2 ),\ldots,\Gamma_n (\Delta_n,Y_n,W_n )$, defined pointwise as
\begin{align*}
    \Gamma_n (\delta,y,w)=  \frac{\delta- \mu_n (y,w)}{g_n (y,w)}+\frac{1}{n}\sum_{j=1}^n \mu_n (y,W_j )\ ,
\end{align*}
are then obtained and regressed against $Y_1,Y_2,\ldots,Y_n$ over the domain $[t_0,t_1]$ using isotonic regression. Technical details are provided in Section \ref{sec:CIR details} of the Supplementary Material, including how it relates to the isotonization procedure of \citet{vandervaart2006estimating} in the absence of survey nonresponse. We denote the resulting estimator of $\psi(t)$ as $\psi_n(t)$. Use of these pseudo-outcomes rather than the simple event indicators accounts for possible dependence between $\Delta_*(t)$ and $Y_*$, as long as this dependence is explained entirely by $W$. 

Roughly speaking, if at least one of $\mu$ and $g$ is estimated consistently, then $\psi_n (t)$ is itself consistent for $\psi(t)$, and if both $\mu$ and $g$ are estimated consistently at fast enough rates, $n^{\frac{1}{3}} \{\psi_n (t)-\psi(t)\}$ tends in distribution to a random variable with a standard Chernoff distribution \citep{groeneboom2001computing} scaled by a constant $\tau(t)$. This well-characterized large-sample distribution enables the construction of confidence intervals with correct coverage. The scale factor $\tau(t)$ of the limiting distribution, whose explicit form is given in Section \ref{sec:CIR details} of the Supplementary Material, depends on $\mu$ and $g$, as well as on the derivative of the survival curve at $t$. To estimate $\tau(t)$, we require an estimate of this derivative, which can be obtained by applying a smoothing procedure to the estimated survival curve $\psi_n(t)$ and then using numerical differentiation. Once the scale factor has been estimated, with estimate denoted by $\tau_n(t)$, confidence intervals with nominal pointwise coverage $1-\alpha$ can be constructed as 
\begin{align*}
    \left(\psi_n(t) - q_{1-\alpha/2}\tau_n(t)n^{-\frac{1}{3}}, \psi_n(t) + q_{1-\alpha/2}\tau_n(t)n^{-\frac{1}{3}}\right),
\end{align*}
where $q_\beta$ denotes the $\beta$--quantile of the standard Chernoff distribution. A key strength of the CIR approach is that it allows for flexible estimation of $\mu$ and $g$ --- using machine learning, for example --- without sacrificing statistical inference. This reduces the risk of systematic bias due to the use of misspecified parametric models. In Section \ref{sec:sims} of the Supplementary Material, we present results of a simulation study illustrating the relative insensitivity of our proposed procedure to the choice of nuisance estimators.

We note that performing the CIR procedure using data from survey respondents alone --- in other words, performing a complete case analysis --- could result in bias, since whether a participant is considered a respondent is determined by their value of $Y_*$, which may depend on $W$ and thus be informative for $\Delta_*$. Symbolically, the estimand in a complete case analysis would be $1 - E\{E(\Delta \midd W, Y = t) \midd Y < c_0\}$ rather than $1 - E\{E(\Delta \midd W, Y = t)\}$. Therefore, fully removing nonrespondents --- that is, participants with $Y_* \geq c_0$ --- may induce selection bias, as the distribution of $W$ may differ among survey respondents and nonrespondents. To emphasize the difference between the complete case CIR and our proposed method, which includes all survey recipients, we refer to the latter as \textit{extended CIR}. 

Even with a rich set of baseline covariates $W$, there may be concern that conditional independence of $\Delta_*(t)$ and $Y_*$ is unlikely to hold. In Section \ref{sec:sensitivity analysis} of the Supplementary Material, we detail a sensitivity analysis procedure designed to assess how deviations from conditional independence affect the estimates of $\psi(t)$ produced by the extended CIR procedure. The sensitivity analysis is based on an assumed copula model that encodes dependence between $\Delta_*(t)$ and $Y_*$ within covariate strata \citep{genest1987frank}. By varying the dependence parameter of the copula model, an investigator can observe how sensitive the CIR estimates are to violations of conditional independence. We note that the dependence parameter cannot be estimated from the observed data and must be specified by the investigator in order to estimate $\psi(t)$ under dependence within covariate strata.

It is also of interest to understand the association between baseline covariates $W$ and symptom duration $T$ in order to identify potential predictors of delayed symptom resolution. In Section \ref{sec:cox methods} of the Supplementary Material, we review regression models that have been adapted for use with current status data and discuss the difficulties that arise in making inference using these models. Focusing on the Cox model, we conduct a simulation study to evaluate the performance of the nonparametric bootstrap for making inference, finding that Wald-type bootstrap confidence intervals based on a normal approximation demonstrate good performance in modest sample sizes. The Cox model requires no special adaptation to account for nonresponse, as long as the regression model is correctly specified and includes all covariates that inform the response mechanism \citep{little2019statistical}.

\section*{Analysis of HCT data}

\subsection*{Study setting and participants}
The HCT study was conducted at UW, a large public university in Washington state with a population composed of approximately 60,000 students and 32,000 faculty and staff. Beginning in autumn 2020, the HCT program provided free, voluntary PCR testing for SARS-CoV-2 and symptom surveillance for university affiliates. Individuals were sent daily messages via text or email to ask about new symptoms, exposures, and high-risk behaviors, any of which would trigger an invitation for PCR testing. Study participants were also able to walk-in for testing at any time, and for any reason, without an invitation. Once antigen testing became widely available in March $2022$, individuals were asked on their daily messages to report any antigen test results. The inclusion criteria for HCT eligibility are described in Section \ref{sec:data analysis supp} of the Supplementary Material. Informed consent was signed electronically at the time of enrollment.  

For this analysis, we used data collected from December $2021$ to June $2023$. During this time period, participants who had a positive PCR test or who reported a positive rapid antigen test for SARS-CoV-2 were invited to participate in a follow-up survey to assess for ongoing symptoms of COVID-19 a minimum of four weeks after the positive test.  

\subsection*{Data collection}
All data were collected electronically by sending participants an email or text notification with a link to a questionnaire. 
The surveys included: 
\begin{enumerate}
    \item[(a)] \textit{Enrollment questionnaire and quarterly update survey}: The enrollment questionnaire included demographics, COVID-19 vaccination information, and current pre-existing medical conditions, which were captured by selecting zero or more options from the list given in Table \ref{tab:comorbidities}. A quarterly survey was used to update eligibility and vaccination status.

\begin{table}
	\centering
		\begin{tabular}{lll}  \toprule
            Category &Pre-existing conditions \\ \midrule 
            Cardiometabolic & Diabetes, heart disease, high blood pressure& \\
            Respiratory & Asthma/reactive airway disease, chronic bronchitis, COPD/emphysema&\\
            Immunological & Cancer, immunosuppression \\
            Seasonal allergies & ---\\
\bottomrule
		\end{tabular}
		\caption{Pre-existing medical conditions on which baseline information was collected.}
		\label{tab:comorbidities}
\end{table} 
    
    \item[(b)] \textit{Daily attestation survey and symptom attestation survey at the time of SARS-CoV-2 test}: Participants were prompted to fill out a daily attestation survey that collected information on date of known exposures to COVID-19, and date(s) and results of rapid antigen test(s) performed at home. New or worsening symptoms in the past $24$ hours were captured by selecting zero or more items from the list given in Table \ref{tab:symptoms}. 
    
    \begin{table}
	\centering
		\begin{tabular}{lp{0.7\linewidth}}  \toprule
            Category & Symptoms \\ \midrule 
            Upper respiratory &  Cough, ear pain/discharge, eye pain, loss of smell or taste, rash, runny/stuffy nose, sore/itchy/scratchy throat, trouble breathing\\
            Systemic & Chills or shivering, feeling feverish, headache, muscle or body aches, sweats\\
            Gastrointestinal & Diarrhea, nausea/vomiting \\
            Fatigue & --- \\
\bottomrule
		\end{tabular}
		\caption{List of symptoms during the acute infection phase.}
		\label{tab:symptoms}
\end{table} 
    
    \item[(c)] \textit{Long COVID survey}: Starting in December $2021$, all participants who tested positive for SARS-CoV-2 by PCR testing or those who attested to having had a positive rapid antigen test were contacted a minimum of $28$ days after their first positive test date and invited to participate in the long COVID survey. Each participant was allowed to fill out the survey only once per academic year. For participants with multiple responses, only the first is included in this analysis. The survey included the question ``Are you still experiencing symptoms related to your COVID-19 illness (e.g., brain fog, fatigue, headaches, etc.)?'' Each participant was free to choose when to complete the survey. Data collection was terminated on June $23$, $2023$. 
\end{enumerate}

\subsection*{Risk factors}
For the statistical analyses, a subset of baseline covariates collected on participants was used. Demographic risk factors included self-reported sex (male or female) and age (categorized as $25$ years and under, $26$ to $40$ years, $41$ to $60$ years, and over $60$ years). Baseline comorbidities were categorized as cardiometabolic, respiratory, immunological, and/or seasonal allergies (see Table \ref{tab:comorbidities}). Symptoms at baseline were categorized as upper respiratory symptoms, systemic symptoms, gastrointestinal symptoms, and/or fatigue (see Table \ref{tab:symptoms}). Baseline comorbidities and symptoms were included in analyses as binary (presence/absence). As a proxy for viral load, participants were categorized according to the manner in which they tested positive for SARS-CoV-2 (rapid antigen test versus PCR), and among those who tested positive via PCR, according to the cycle threshold (Ct) value. Higher Ct values correspond to a lower viral load. The categories used were rapid antigen test, PCR with Ct below $30$, and PCR with Ct greater than $30$. Participants were also categorized according to the time elapsed between most recent vaccination and positive test: less than three months, between three and six months, and more than six months. We restricted our analysis to vaccinated participants because there were few unvaccinated participants --- less than $2$\% of the study cohort --- and because vaccination was required by UW policy during the data collection period, rendering unvaccinated individuals possibly different from most of the UW population in important ways beyond vaccination status. 

\subsection*{Implementation}
We implemented the extended CIR method using flexible machine learning tools to estimate $\mu$ and $g$. To estimate $\mu$, we used the Super Learner \citep{van2007super, rose2013mortality}, a form of stacked regression, with the optimal linear combination of learners selected to minimize $10$-fold cross-validated mean squared error --- see Section \ref{sec:data analysis supp} of the Supplementary Material for the algorithm library. To estimate $\pi$, we used the \texttt{haldensify} software package \citep{hejazi2022haldensify} with $10$-fold cross-validation used to select optimal tuning parameters. The derivative of the survival curve, which is required for confidence interval construction, was estimated by numerical differentiation of a monotone Hermite spline approximation of the estimated survival curve. Confidence intervals with nominal $95$\% pointwise coverage were constructed using the 97.5th quantile of the Chernoff distribution. 

We aimed to estimate the survival function for symptom resolution between $t_0=30$ and $t_1=90$ days after the positive test. We therefore chose the maximum follow-up time to be $c_0 = 120$; participants with positive tests after February $24$, $2023$ were excluded to ensure all individuals in the analysis cohort had at least $120$ days of follow-up. On one hand, large values of $c_0$ result in more individuals excluded from the study period but make $t_1$ further away from the largest possible observed survey response times, where the statistical performance of isotonic regression and related procedures is known to degrade. On the other hand, small values of $c_0$ result in fewer individuals excluded from the study period but make $t_1$ closer to the largest possible observed survey response times. Our choice of $c_0$ was meant to achieve a balance between these competing considerations, and resulted in $t_1$ lying at approximately the $97.5$th percentile of observed survey response times; such choice performed well in a simulation study --- see Section \ref{sec:sims} of the Supplementary Material for details.

We also performed a sensitivity analysis using the procedure detailed in Section \ref{sec:sensitivity analysis} of the Supplementary Material.  We varied the copula association parameter to encode different degrees of conditional dependence between $T$ and $Y_*$ given $W$, corresponding to values of Kendall's rank correlation coefficient ranging from -0.25 to 0.25. We used the same estimated nuisance functions as in the main analysis so that any observed variation in results can be directly ascribed to deviations from the conditionally uninformative response time assumption.

To investigate the association between baseline risk factors and time from infection to symptom resolution, we fit a Cox proportional hazards regression  model accounting for current status sampling using the \texttt{icenReg} software package \citep{anderson2017icenreg}. Main terms were included in the model for each baseline covariate described above. Standard error estimates were calculated using a nonparametric bootstrap with $1000$ bootstrap replicates, and used to construct $95$\% Wald-type confidence intervals and perform two-sided tests of the null regression coefficient hypothesis at a $0.05$ level. For categorical risk factors, multivariate Wald tests were used to test the joint null hypothesis that all of the corresponding regression coefficients were zero. 
	
 \subsection*{Results} 
After the removal of $285$ individuals due to missing covariate information, the analysis cohort consisted of $3489$ HCT study participants with positive SARS-CoV-2 tests prior to February $24$, $2023$. Of these, $1434$ ($41.1$\%) responded to the survey. Characteristics of the analysis cohort are summarized in Table \ref{tab:descriptives}. The median age was $23$ years (IQR: $20$--$36$), and $64$\% of the participants reported female sex. Slightly over $33$\% of participants reported at least one baseline comorbidity, and roughly $52$\% were symptomatic at the time of testing. Most participants were not recently vaccinated. Survey respondents tended to be older than nonrespondents, with markedly higher prevalence of symptoms during the acute infection phase. 

\begin{table}
	\centering
		\begin{tabular}{lccc}  \toprule
  & Respondents & Nonrespondents&Overall analysis cohort \\
  & $n = 1434$ & $n = 2055$ & $n = 3489$\\ \midrule
 Age, years &&\\
  \hspace{0.25in}25 and under &597 (41.6\%)&1436 (69.9\%)&2033 (58.3\%)\\
  \hspace{0.25in}26--40  &362 (25.2\%)&385 (18.7\%)&747 (21.4\%)\\
  \hspace{0.25in}41--60 &376 (26.2\%)&179 (8.7\%)&555 (15.9\%)\\
  \hspace{0.25in}61 and over &99 (6.9\%)&55 (2.7\%)&154 (4.4\%)\\ 
  Male &495 (34.5\%)&762 (37.1\%)&1257 (36.0\%)\\
  Symptoms, upper respiratory &1028 (71.7\%)&709 (34.5\%)&1737 (49.8\%)\\
  Symptoms, systemic&715 (49.9\%)&480 (23.4\%)&1195 (34.3\%)\\
    Symptoms, gastrointestinal &122 (8.5\%)&113 (5.5\%)&235 (6.7\%)\\
   Symptoms, fatigue & 466 (32.5\%)&331 (16.1\%)&797 (22.8\%)\\
  Comorbidities, cardiometabolic & 96 (6.7\%)&50 (2.4\%)&146 (4.2\%)\\
   Comorbidities, respiratory & 147 (10.3\%)&196 (9.5\%)&343 (9.8\%)\\
   Comorbidities, immunological & 29 (2.0\%)&25 (1.2\%)&54 (1.5\%)\\
   Comorbidities, seasonal allergy & 434 (30.3\%) &445 (21.7\%)&879 (25.2\%)\\
  Time since last vaccination & &\\
  \hspace{0.25in} Under 3 months &309 (21.5\%)&265 (12.9\%)&574 (16.5\%)\\
  \hspace{0.25in} 3--6 months & 202 (14.1\%)&314 (15.3\%)&516 (14.8\%)\\
  \hspace{0.25in} More than 6 months & 923 (64.4\%)&1476 (71.8\%)&2399 (68.8\%)\\
  Viral load & \\
  \hspace{0.25in} PCR, over 30 Ct & 273 (19.0\%)&452 (22.0\%)&725 (20.8\%) \\
  \hspace{0.25in} PCR, under 30 Ct & 545 (38.0\%)&924 (45.0\%)&1469 (42.1\%)\\
  \hspace{0.25in} Rapid antigen test &616 (43.0\%)&679 (33.0\%)&1295 (37.1\%)\\

\bottomrule
		\end{tabular}
		\caption{Baseline characteristics of the analysis cohort.}
		\label{tab:descriptives}
\end{table} 

\begin{figure}
\centering
		\includegraphics[width=0.9\linewidth]{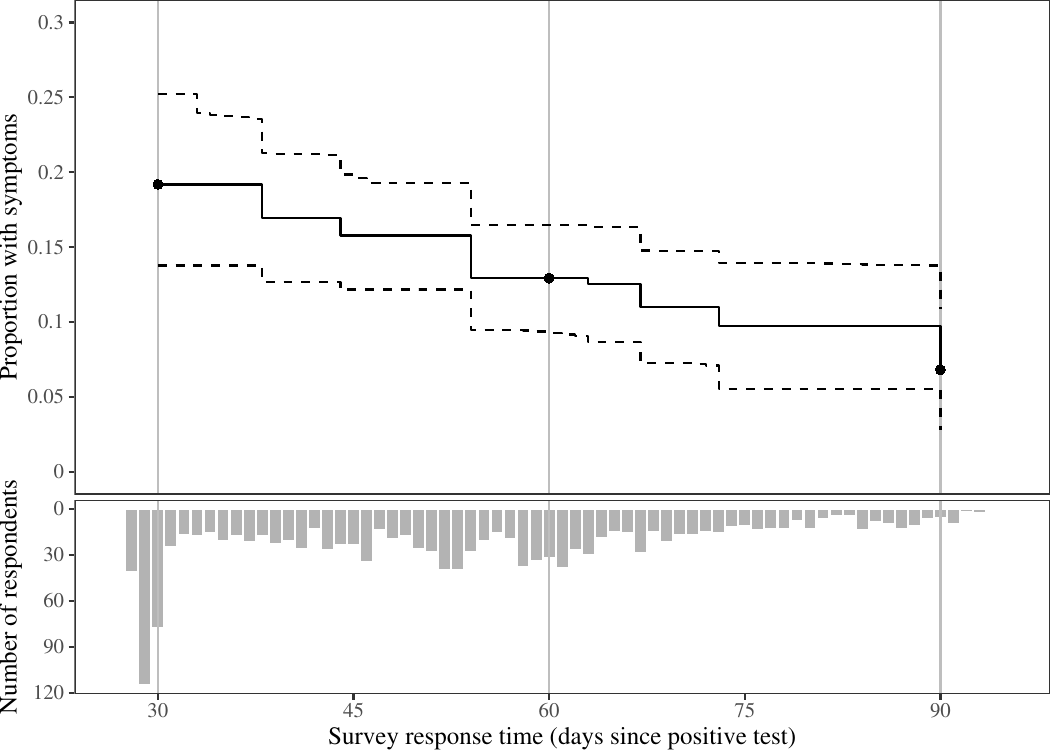}
	\caption{Standardized survival curve for time from positive test until symptom resolution, estimated using the proposed extended causal isotonic regression method (top panel), along with the distribution of survey response times between $28$ and $95$ days (bottom panel). The solid black line shows the estimated proportion with ongoing symptoms at times from $30$ days to $90$ days after the positive test. The black points show the estimated proportion with ongoing symptoms at $30$, $60$, and $90$ days. The dashed lines represent a pointwise 95\% confidence interval. The gray vertical lines denote $30$, $60$, and $90$ days.}
	\label{fig:survival curve}
\end{figure}

Figure \ref{fig:survival curve} shows the estimated proportion of participants with persistent COVID-19 symptoms at times from 30 to 90 days following the positive test, along with the distribution of survey response times. From the extended CIR method, estimated symptom persistence proportions at $30$, $60$, and $90$ days after the positive test were $19.2$\% (95\% CI: $13.7$--$25.2$), $12.9$\% ($9.4$--$16.5$), and $6.8$\% ($2.8$--$10.9$), respectively. 

\begin{figure}
\centering
		\includegraphics[width=0.9\linewidth]{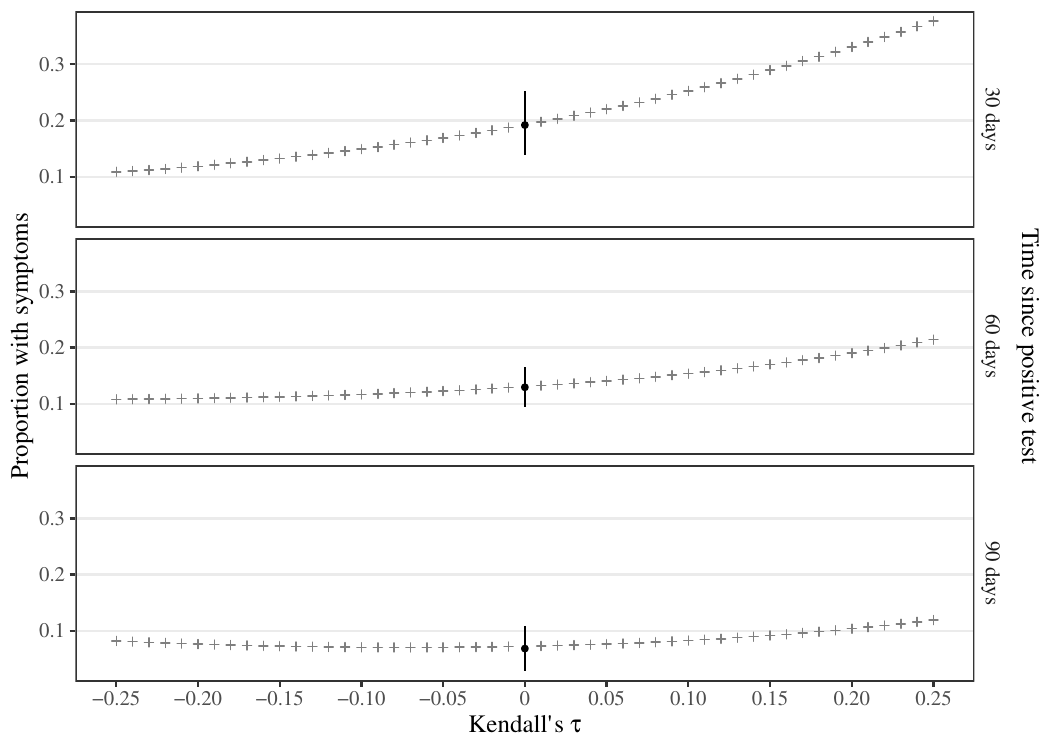}
	\caption{Sensitivity analysis for deviation from conditionally uninformative response time assumption. From top to bottom, the rows correspond to the estimated proportion with ongoing symptoms at $30$, $60$, and $90$ days after the positive test. The $x$-axis represents the value of Kendall's rank correlation coefficient $\tau$ used to encode conditional dependence in each sensitivity analysis. At $\tau = 0$, in black, we give the estimate and confidence interval obtained from the proposed extended CIR procedure assuming conditional independence (primary analysis). Gray crosses represent estimates corresponding to different values of $\tau$.}
	\label{fig:copula survival curve}
\end{figure}

Figure \ref{fig:copula survival curve} shows the results of the sensitivity analysis performed. The extended CIR estimates are substantially more sensitive to violations of conditional independence at the earliest time point, 30 days after the positive test, compared to 60 and 90 days. At 30 days, under negative dependence, meaning that those with longer symptom resolution times tend to respond more quickly to the survey, the estimated proportion with persistent symptoms is lower than in the primary analysis. Under positive dependence, the opposite pattern emerges and also reveals greater sensitivity. At 60 and 90 days, negative dependence has little discernible impact on the estimated proportion with ongoing symptoms, whereas positive dependence results in a moderately higher estimated proportion, particularly at 60 days.

Table \ref{tab:cox results} contains the results of the Cox proportional hazards regression, with hazard ratios larger than $1$ indicating faster symptom resolution. Participants identifying as male tended to experience faster symptom resolution, with an estimated hazard ratio for resolution of $1.21$ ($1.04$--$1.42$, $P = 0.014$). Conversely, participants reporting fatigue at baseline tended to have longer lasting symptoms, with an estimated hazard ratio of $0.77$ ($0.65$--$0.93$, $P = 0.005$). In terms of viral load, participants with Ct values under $30$ and participants who tested positive via rapid antigen tests had more persistent symptoms compared to participants with Ct values over $30$, with estimated hazard ratios of $0.78$ ($0.64$--$0.96$) and $0.76$ ($0.60$--$0.96$), respectively ($P = 0.034$). 

\begin{table}
	\centering
		\begin{tabular}{lccc}  \toprule
  Risk factor & Hazard ratio & 95\% confidence interval & $P$-value\\ \midrule 
  Age, years &&&0.610\\
  \hspace{0.25in}25 and under &(ref)&(ref)&(ref)\\
  \hspace{0.25in}26--40  &0.923&(0.767, 1.111)&---\\
  \hspace{0.25in}41--60 &0.914 &(0.749, 1.115)&---\\
  \hspace{0.25in}61 and over &1.075 &(0.777, 1.487)&---\\ 
  Male &\textbf{1.214}&\textbf{(1.040, 1.417)}&\textbf{0.014}\\
  Symptoms, upper respiratory &1.085&(0.891, 1.322)&0.418\\
   Symptoms, systemic &0.964&(0.802, 1.160)&0.699 \\
  Symptoms, gastrointestinal &0.885&(0.674, 1.164)&0.383\\
Symptoms, fatigue &\textbf{0.774}&\textbf{(0.646, 0.927)}&\textbf{0.005}\\
 Comorbidities, cardiometabolic &0.809&(0.602, 1.087)&0.159 \\
  Comorbidities, respiratory &0.869&(0.690, 1.095)&0.234 \\
  Comorbidities, immunological &0.840&(0.490, 1.439)&0.526 \\
 Comorbidities, seasonal allergy &0.864&(0.741, 1.006)&0.060 \\
  Time since last vaccination &&&0.798 \\
  \hspace{0.25in} Under 3 months &(ref)&(ref)&(ref) \\
  \hspace{0.25in} 3--6 months &0.946&(0.746, 1.200)&--- \\
  \hspace{0.25in} More than 6 months &0.931&(0.752, 1.151)&--- \\
  Viral load &&&\textbf{0.034} \\
  \hspace{0.25in} PCR, over 30 Ct &(ref)&(ref)&(ref) \\
  \hspace{0.25in} PCR, under 30 Ct &\textbf{0.782}&\textbf{(0.636, 0.962)}&--- \\
  \hspace{0.25in} Rapid antigen test &\textbf{0.755}&\textbf{(0.596, 0.956)}&---\\
\bottomrule
		\end{tabular}
		\caption{Results of Cox proportional hazards regression for time to symptom resolution. Boldface font indicates statistical significance at a $0.05$ level. For categorical variables, $P$-values correspond to a multivariate test of null regression coefficients. Ct: cycle threshold.}
		\label{tab:cox results}
\end{table} 

\section*{Discussion}

In this article, we presented modern methodology for the analysis of current status data arising in epidemiological studies, with a particular focus on the novel extension and application of causal isotonic regression to this setting. When it is of interest to estimate the distribution of disease duration under current status sampling, CIR offers a flexible approach accommodating the use of off-the-shelf machine learning tools. Unlike the commonly used nonparametric maximum likelihood estimator, the validity of CIR relies only on conditional independence of survey response times and event indicators within strata of measured baseline covariates, thereby weakening the assumptions under which valid results will be obtained. In addition, we have shown how CIR can be implemented to properly account for survey nonresponse. 

We demonstrated the use of CIR, along with existing survival analysis regression methods, to analyze current status data from the long COVID survey study conducted at UW between December $2021$ and June $2023$. Participants were queried at least $28$ days after testing positive, and even for those queried exactly at $28$ days, many did not respond until days or weeks later. This variation in response times motivates the use of the statistical methods described in this paper. Using the long COVID survey data, we estimated the prevalence of persistent COVID symptoms in the UW population to be $19$\% at $30$ days after a positive test, dropping to $7$\% at $90$ days. 

A retrospective study in a comparable university population found that $36$\% of survey respondents reported symptoms lasting at least 28 days from the end of the recommended $10$-day isolation period \citep{landry2023postacute}. Our results provide further evidence that, even in a predominantly young and vaccinated population, a substantial proportion of infected individuals are affected by persistent COVID-19 symptoms. Notably, because participants were asked only about persistent symptoms and not about new symptoms arising in the post-acute phase, our analysis may still represent an underestimate of the frequency of PASC, which may also be characterized by symptoms that differ from initial acute symptoms and arise after their resolution.

We also investigated potential risk factors for persistent symptoms using Cox proportional hazards regression, and found evidence that female sex, fatigue during the acute phase, and higher semi-quantitative viral load at time of testing are associated with slower resolution of symptoms. This is in line with previous studies that have identified female sex \citep{su2022multiple}, severe fatigue during the acute infection phase \citep{shen2023covid}, and higher peak semi-quantitative viral load in the four weeks following diagnosis \citep{lu2023early} as potential risk factors for PASC. 

The validity of extended CIR hinges on the untestable assumption that survey response times and event indicators are conditionally independent given measured baseline covariates, which is more likely to be satisfied through the inclusion of potential risk factors as covariates in the analysis. If the survey response time depends on symptom duration even within strata defined by baseline covariates, the extended CIR procedure could be susceptible to bias. We derived one approach for formal sensitivity analysis in this setting, which allows investigators to examine changes in the estimated survival function as the dependence between response times and symptom resolution times, given measured covariates, deviates from zero. The development of additional sensitivity analysis methods based on alternative deviations is an important avenue of future research. Further work is also needed to better understand the practical implications of the poor boundary performance of CIR in order to provide a principled approach to choose appropriate values of $t_0, t_1$ and $c_0$. 

\singlespacing
\section*{Acknowledgements}

We would like to thank the participants of the Husky Coronavirus Testing study. This work was supported by the National Science Foundation Graduate Research Fellowship Program under Grant No. DGE-2140004, the National Heart, Lung, and Blood Institute under grant R01-HL137808, and the CARES act under University of Washington Grant 624611.

\section*{Software and code}

The extended CIR method is implemented in the \texttt{survML} \texttt{R} package, available at \url{https://github.com/cwolock/survML}. Code to reproduce all results is available online at \url{https://github.com/cwolock/currstat_CIR_supplementary}. 

\section*{Conflict of interest statement}
HYC has consulted for Bill and Melinda Gates Foundation and Ellume, and has served on advisory boards for Vir, Merck and Abbvie. She has received research funding from Gates Ventures, and support and reagents from Ellume and Cepheid outside of the submitted work. 

\newpage 
\singlespacing
\bibliographystyle{apalike}
\bibliography{refs}

\newpage
\doublespacing
\titleformat{\section}{\scshape\bfseries}{S\arabic{section}.}{1em}{}
\titleformat{\subsection}{\normalfont\itshape}{S\arabic{section}.\arabic{subsection}}{1em}{}
\renewcommand{\thefigure}{S\arabic{figure}}
\setcounter{figure}{0}
\renewcommand{\thetable}{S\arabic{table}}
\setcounter{table}{0}
\renewcommand{\thesection}{S\arabic{section}}
\begin{center}
    {\large \textbf{Supplementary Material }}
\end{center}

\section{Causal isotonic regression for current status data}\label{sec:CIR details}

\subsection{Data structure and target of estimation}

We denote by $T$ the unobserved symptom duration after testing positive for SARS-CoV-2. For each study participant, we observe a vector $W \in \mathbb{R}^p$ of baseline covariates. In the context of current status sampling, the \textit{ideal data} unit --- that is, the data unit that would be observed given unlimited follow-up --- consists of $W$, the survey response time $Y_* \in [b_0, \infty)$, and the event indicator $\Delta_* \in \{0,1\}$. We denote by $c_0$ the maximum follow-up time of the study. In reality, we observe $Y := \min(Y_*, c_0)$ and $\Delta := \I(Y_* < c_0)\Delta_*$. For individuals who have not responded to the survey by time $c_0$, we do not observe the event indicator. Thus, the \textit{observed data} unit is $Z := (W, Y, \Delta)$. In this setting, survey non-response corresponds to having failed to respond to the survey prior to time $c_0$. 
The available data consist of $n$ independent observations $Z_1, Z_2, \ldots, Z_n$ draw from distribution $P_0$, which is assumed to lie in a nonparametric model $\mathcal{M}$. Here and after, we use the subscript 0 to denote a functional of the true data-generating distribution $P_0$, e.g., $E_0\{f(W)\} := E_{P_0}\{f(W)\}$. For simplicity of notation, in what follows we consider estimation of the cumulative distribution function $t \mapsto \psi_0(t) := P_0(T \leq t)$. (In the main text, $\psi$ was used to denote the survival function, which is simply equal to one minus the distribution function). 

For a generic distribution $P \in \mathcal{M}$, we define the mapping
\begin{align}\label{eq:ID}
    \theta_P: y_0 \mapsto E_P\left\{E_P\left(\Delta \midd W, Y = y_0\right)\right\}. 
\end{align}
For all $y_0 \in (b_0, c_0)$, the target parameter $\psi_0(y_0)$ is identified by the observed data as $\theta_0(y_0)$. If $y_0 < c_0$, then the event $\{Y_* = y_0\}$ is equivalent to the event $\{Y = y_0\}$, and so this parameter is equal to the distribution function of $T$ evaluated at $y_0$ under the conditional independence assumption described in the main text. 

We note that the outer expectation in (\ref{eq:ID}) is taken with respect to the marginal distribution of $W$ under sampling from $P_0$, as opposed to the distribution of $W$ given $Y < c_0$. This suggests that performing a standard causal isotonic regression (CIR) procedure among survey respondents would not in general yield a consistent estimator, as the marginal distribution of $W$ and the conditional distribution of $W$ given $Y < c_0$ are not necessarily equal unless $W$ and $Y$ are independent. 

\subsection{Estimation and inference}\label{sec:estimation and inference supplement}

Without making strong modeling assumptions, the target regression function $\theta_0(y_0)$ is not pathwise differentiable and is therefore not estimable at $n^{-\frac{1}{2}}$--rate. However, it is known to be monotone as a function of $y_0$ and can be viewed as the target of an isotonic regression procedure. Using isotonic regression to estimate $\theta_0(y_0)$ allows for the application of established statistical theory \citep{westling2020unified} to characterize the large-sample properties of the resulting estimator. We outline this isotonic regression procedure here. 

For a generic $P \in \mathcal{M}$, we first define the outcome regression function $\mu_P: (y,w) \mapsto E_P(\Delta \midd W = w, Y = y)$ and density ratio $g_P: (y,w) \mapsto \pi_P(y \midd w)/f_P(y)$ with $\pi_P$ the conditional density of $Y$ given $W$ and $f_P$ the marginal density of $Y$. These nuisance functions play a key role in accounting for the possible dependence between $\Delta_*$ and $Y_*$. We use $F_P$ to denote the marginal distribution function of $Y$. We use $F_P$ as a domain transformation function in the CIR procedure. While the CIR procedure could be performed without the use of this domain transformation, its use results in a procedure that is invariant to strictly increasing transformations of $Y$ \citep{westling2020causal}. In other words, changing the scale on which response times are measured will have no impact on the estimated survival function. 

We next define $\psi_P := \theta_P \circ F_P^{-1}$ and corresponding primitive function $\Psi_P(t) := \int_0^t \psi_P(u)du$. The causal isotonic regression (CIR) procedure entails estimation of the parameter $\Gamma_P := \Psi_P \circ F_P$, which takes the form
\begin{align*}
    \Gamma_{P}: y_0 \mapsto \iint \I(u \leq y_0)\mu_P(u,w)F_P(du)Q_P(dw)
\end{align*}
with $Q_P$ the marginal distribution function of $W$ under sampling from $P$. Unlike $\theta_0(y_0)$, the parameter $\Gamma_0(y_0)$ is pathwise differentiable relative to the nonparametric model $\mathcal{M}$ and can therefore be estimated at $n^{-\frac{1}{2}}$--rate  \citep{pfanzagl1990estimation}. Per the results of \citet{westling2020causal}, the efficient influence function of $P \mapsto \Gamma_P(y_0)$ relative to $\mathcal{M}$ evaluated at $P_0$ is given by
\begin{align*}
    &D^*_{0, y_0}: (\delta, y, w) \mapsto \\
    &\hspace{1.5cm}\I(y \leq y_0)\left\{\frac{\delta - \mu_0(y,w)}{g_0(y,w)}\right\} + \int \I(u \leq y_0)\mu_0(u,w)F_0(du) 
    + \I(y \leq y_0)\theta_0(y) - 2\Gamma_0(y_0)\ .
\end{align*}
Knowledge of this efficient influence functions enables standard debiasing approaches, such as the one-step correction \citep{Pfanzagl1982}, to be used in estimation of $\Gamma_0(y_0)$. This yields a debiased, asymptotically linear estimator $\Gamma_n(y_0)$ of $\Gamma_0(y_0)$, which is used in the isotonic regression procedure to construct an estimator of $\theta_0(y_0)$. Our proposed procedure is as follows. 
\begin{enumerate}
    \item[(1)] Select domain bounds $t_0$ and $t_1$ satisfying $[t_0,t_1] \subset (b_0,c_0)$. 
    \item[(2)] Construct an estimate $P_n$ of $P_0$ using estimators $(\mu_n, g_n, F_n, Q_n)$ of $(\mu_0, g_0, F_0, Q_0)$. For example, $\mu_n$ and $g_n$ can be constructed using machine learning algorithms, while $F_n$ and $Q_n$ can be constructed using the empirical distributions of $Y$ and $W$, respectively, based on the observed data $Z_1, Z_2, \ldots, Z_n$. 
    \item[(3)] Let $\mathcal{R}_n$ denote the set of unique values of $Y_1, Y_2, \ldots, Y_n$. For each $y \in \mathcal{R}_n$ satisfying $t_0 \leq y \leq t_1$, compute the one-step debiased estimator $\Gamma_n(y)$ of $\Gamma_0(y)$, given by $\Gamma_n(y) := \Gamma_{P_n}(y_0) + \frac{1}{n}\sum_{i=1}^{n}D^*_{P_n, y}(Z_i)$.
    \item[(4)] Compute the greatest convex minorant (GCM) of the set of points
    \begin{align*}
        \{(0,0)\} \cup \{(F_n(Y_i), \Gamma_n(Y_i)): i = 1, 2, \ldots, n, t_0 \leq Y_i \leq t_1\}
    \end{align*}
    Set $\psi_n$ to be the left derivative of the GCM. 
    \item[(5)] Compute $\theta_n(y_0) := \psi_n\{F_n(y_0)\}$ as an estimator of $\theta_0(y_0)$. 
\end{enumerate}
We note that, were we not to apply the domain transformation $F_P$, we could perform a similar procedure using the primitive parameter $y_0 \mapsto \iint \I(u \leq y_0)\mu_P(u,w)Q_P(dw)$. In the absence of survey nonresponse, this procedure would take the same form as the isotonization method of \citet{vandervaart2006estimating}. However, it is not clear whether this approach can be framed in terms of isotonic regression using pseudo-outcomes, as described in the main text. 

If for every point $y_0$ in the estimation domain $[t_0,t_1]$, either $\mu_n$ converges to $\mu_0$ or $g_n$ converges to $g_0$, Theorem 1 of \citet{westling2020causal} establishes that $\theta_n(y_0)$ converges in probability to $\theta_0(y_0)$ for each $y_0\in[t_0,t_1]$. The conditions used to establish distributional convergence are more complex. Roughly speaking, we require that the product of the convergence rates of $\mu_n$ and $g_n$ to their respective targets is $n^{-\frac{1}{3}}$. If these conditions are satisfied, Theorem 2 of \citet{westling2020causal} establishes that, for any $y_0 \in (b_0,c_0)$ with $F_0(y_0) \in (0,1)$, $n^{\frac{1}{3}}\left\{\theta_n(y_0) - \theta_0(y_0)\right\}$ converges in distribution to $\tau_0(y_0)^{\frac{1}{3}}\mathbb{W}$, where $\mathbb{W}$ follows a standard Chernoff distribution \citep{groeneboom2001computing} and $\tau_0(y_0) := 4\theta_0'(y_0)\kappa_0(y_0)/f_0(y_0)$ with 
\begin{align*}
    \kappa_0(y_0) := \int \frac{E_0[\{\Delta - \mu_0(y_0,w)\}^2 \midd W = w, Y = y_0]}{g_0(y_0 \midd w)}\,Q_0(dw)\ .
\end{align*}
In light of the fact that $\Delta$ is binary, we can rewrite the above display as
\begin{align*}
     \kappa_0(y_0) = \int \frac{\mu_0(y_0,w)\left\{1 - \mu_0(y_0,w)\right\}}{g_0(y_0 \midd w)}\,Q_0(dw)\ .
\end{align*}
Construction of confidence intervals requires estimation of the scale factor $\tau_0(y_0)$. We note that $\kappa_0(y_0)$ can be estimated by simply plugging in available estimates of $\mu_0$ and $g_0$, with the empirical distribution of $W_1, W_2, \ldots, W_n$ used to estimate $Q_0$. To estimate $f_0$, we note that $f_0(y) = \int \pi_0(y \midd w)Q_0(dw)$, which be estimated by $\frac{1}{n}\sum_{i=1}^{n}\pi_n(y \midd W_i)$ for a conditional density estimator $\pi_n$. To estimate $\theta_0'(y_0)$, the derivative of $y \mapsto \theta_0(y)$ evaluated at $y_0$, there are a number of available options. In our implementation, we construct a monotone Hermite spline approximation of $\theta_n$ and perform numerical differentiation. 

\section{Sensitivity analysis}\label{sec:sensitivity analysis}

In practice, there may be concern that conditional independence of $T$ and $Y_*$ given $W$ does not hold, potentially rendering the proposed estimation procedure systematically biased. In this section, we describe an approach to sensitivity analysis with respect to this key assumption. This sensitivity analysis approach employs copulas, which are commonly used in multivariate analysis to model the dependence between two or more variables.

A copula is a multivariate cumulative distribution function, indexed by an association parameter $\alpha$, that describes the joint distribution of two or more random variables, each with a uniform marginal distribution. Sklar's theorem \citep{sklar1959fonctions} states that any multivariate distribution can be expressed in terms of its marginal distributions and a copula. Throughout this section, we use $F$ and $f$ to denote generic distribution and density functions, respectively, with a subscript indicating the distribution and random variable corresponding to the given distribution/density function, e.g., $F_{P; T \midd W}(t \midd w) := P(T \leq t \midd W = w)$.

To begin, suppose that for some copula $C_\alpha$ the joint distribution function of $T$ and $Y$ given $W=w$ can be expressed as $
    F_{P;T, Y \midd W}(t, y \midd w) = C_\alpha \{F_{P;T \midd W}(t \midd w), F_{P;Y \midd W}(y \midd w)\}$
for all $t$, $y$ and $w$. It is not difficult to show then that $
    F_{P; T \midd Y, W}(t \midd y,w)= h_{\alpha}\{F_{P;T \midd W}(t \midd w),F_{P;Y \midd W}(y \midd w)\}$, where  $h_{\alpha}$ is defined as the mapping $(u,v) \mapsto \left.\frac{\partial }{\partial \tilde{v}}C_\alpha(u,\tilde{v})\right\vert_{\tilde{v} = v}$. 
Denoting by $m_\alpha(u,v)$ the value $m$ such that $h_{\alpha}(m,v)=u$, this yields that
$F_{P;T \midd W}(t \midd w) = m_\alpha\{F_{P;T \midd Y, W}(t \midd y,w),F_{P;Y \midd W}(y \midd w)\}$, which implies, in particular, that
\begin{align*}
    F_{P;T \midd W}(y \midd w)\ &=\ m_{\alpha}\{F_{P;T \midd Y, W}(y \midd y,w),F_{P;Y \midd W}(y \midd w)\}\\
    &=\ m_{\alpha}\{\mu_{P}(y, w),F_{P;Y \midd W}(y \midd w)\}\ .
\end{align*}
Defining the mapping $\theta_{P,\alpha}:y_0\mapsto E_P[m_{\alpha}\{\mu_P(y_0,W),F_{P;Y\midd W}(y_0\midd W)\}]$, we find thus that $\theta_{P_0,\alpha}(y_0)$ identifies the target parameter $\psi_0(y_0)$ under the dependence copula model considered. Applying the domain transformation $F_{P;Y}$, we obtain the identified primitive function given, for each $\alpha$, by
\begin{align*}
    \Gamma_{P, \alpha}: y_0 \mapsto \iint \I(u \leq y_0)m_{\alpha }\{\mu_P(u,w),F_{P; Y \midd W}(u \midd w)\}F_{P; Y}(du)F_{P;W}(dw)\ .
\end{align*}
When $\alpha = 0$, this yields exactly the identification $\Gamma_P$ obtained under the assumption that $Y$ and $T$ are conditionally independent given $W$. Our proposed sensitivity analysis approach entails deriving the efficient influence function of this primitive, constructing a debiased one-step estimator, and using this estimator in the CIR procedure outlined in Section \ref{sec:estimation and inference supplement}. In this sense, this sensitivity analysis consists of a strict generalization of the extended CIR procedure developed under conditional independence.

Many commonly used copula functions do not appear to yield a tractable closed form for $m_\alpha$ \citep{bernard2015conditional}, hindering application of the extended CIR procedure. The commonly used Frank copula \citep{genest1987frank} does, however, yield a usable form. The Frank copula can be written as $C_{\alpha}(u,v) = \varphi^{-1}\{\varphi(u) + \varphi(v)\}$ with generator function $\varphi: u \mapsto -\log\left\{\frac{\exp(-\alpha u)-1}{\exp(-\alpha)-1}\right\}$. The association parameter $\alpha$ takes values in $(-\infty, \infty)\setminus \{0\}$, with independence recovered as $\alpha \to 0$. The Kendall's rank correlation coefficient $\tau$ can be expressed explicitly in terms of the Frank copula association parameter $\alpha$ via the formula $\tau = 1 - \frac{4}{\alpha}\left\{1 - \Omega(\alpha)\right\}$, where $\Omega(\alpha) = \frac{1}{\alpha}\int_0^\infty \frac{t}{\exp(t) - 1}\,dt$.

When $C_\alpha$ is taken to be the Frank copula, it is possible to show that
\begin{align*}
  m_{\alpha}(u,v) &= -\frac{1}{\alpha}\log\left\{1 - \frac{u(1 - e^{-\alpha})}{e^{-\alpha v} + u(1 - e^{-\alpha v})}\right\}.
\end{align*}
Here and after, for compactness we revert to the notation used in Section \ref{sec:CIR details}, with $\mu_P(y,w) = E_P(\Delta \midd W = w, Y = y)$, $g_P(y,w) = \pi_P(y \midd  w)/f_P(y)$, $F_P$ the marginal distribution function of $Y$, and $Q_P$ the marginal distribution function of $W$. Furthermore, we use $H_P$ to denote the conditional cumulative distribution function of $Y$ given $W$, that is, $H_P(y,w) := P(Y \leq y \midd W = w)$. We define $\lambda_{P;\alpha}: y \mapsto \int m_{\alpha}\{\mu_P(y,w), F_{P; Y \midd W}(y \midd w)\}F_{P;W}(dw)$. Then, using the delta method for influence functions and defining 
\begin{align*}
    r_\alpha(u,v) := \left\{e^{-\alpha v} + u\left(e^{-\alpha}-e^{-\alpha v}\right)\right\}\left\{e^{-\alpha v} + u\left(1 - e^{-\alpha v}\right)\right\},
\end{align*}
 we find that the nonparametric efficient influence function $D^*_{0, \alpha, y_0}$ of $P \mapsto \Gamma_{P, \alpha}(y_0)$ evaluated at $P_0$ is given pointwise by
\begin{align*}
     (\delta, y, w)\mapsto\ &\frac{\left(1-e^{-\alpha}\right)\I(y \leq y_0)\left\{\delta - \mu_0(y,w)\right\}e^{-\alpha H_0(y,w)}}{\alpha g_0(y, w)r_\alpha\{\mu_0(y,w), H_0(y,w)\}}- 2\int_{u \leq y_0} \lambda_{0;\alpha}(u)F_0(du)\\
    &+\I(y \leq y_0)\lambda_{0; \alpha}(y)+ \int_{u \leq y_0} m_\alpha\{\mu_0(u,w), H_0(u,w)\}F_0(du) \\
    &+ \left(1 - e^{-\alpha}\right)\int_{u \leq y_0}\frac{\left\{\I(y \leq u) - H_0(u,w)\right\}\mu_0(u,w)e^{-\alpha H_0(u,w)}\left\{1 - \mu_0(u,w)\right\}}{r_\alpha\{\mu_0(u,w), H_0(u,w)\}}F_0(du)\ .
\end{align*}
We note that $\lim_{\alpha \to 0}m_\alpha(u,v) = u$, which can be used to verify that the above expression simplifies to the efficient influence function for $P \mapsto \Gamma_P(y_0)$ evaluated at $P_0$ as $\alpha \to 0$. 
Using the empirical distributions $F_n$ and $Q_n$ to estimate $F_0$ and $Q_0$, respectively, and with generic estimators $\mu_n$, $g_n$ and $H_n$ of $\mu_0$, $g_0$ and $H_0$, respectively, the one-step estimator takes the form
\begin{align*}
&\frac{1}{n}\sum_{i=1}^{n}\frac{\I(Y_i \leq y_0)\left\{\Delta_i - \mu_n(Y_i,W_i)\right\}e^{-\alpha H_n(Y_i,W_i)}(1-e^{-\alpha})}{\alpha g_n(Y_i, W_i)r_\alpha\{\mu_n(Y_i,W_i), H_n(Y_i,W_i)\}} + \frac{1}{n}\sum_{i=1}^{n}\I(Y_i \leq y_0)\lambda_{n,\alpha}(Y_i)\\
&+ \frac{1}{n^2}\sum_{i=1}^{n}\sum_{j=1}^{n}\frac{\I(Y_j \leq y_0)\left\{\I(Y_i \leq Y_j) - H_n(Y_j,W_i)\right\}\mu_n(Y_j,W_i)e^{-\alpha H_0(Y_j,W_i)}\left\{1 - \mu_n(Y_j,W_i)\right\}\left(1 - e^{-\alpha}\right)}{r_\alpha\{\mu_n(Y_j,W_i), H_n(Y_j,W_i)\}}\ .
\end{align*}
To perform the sensitivity analysis, we use the above one-step estimator in the extended CIR procedure over a range of values of the association parameter $\alpha$.

As noted in the main text, the association parameter $\alpha$ is not identified by the observed data and therefore cannot be estimated. To show this, we construct two distributions $P_1$ and $P_2$ for the ideal data unit $(T,Y,W)$, with respective association parameters $\alpha_1$ and $\alpha_2$ with $\alpha_1\neq \alpha_2$, that imply the same distribution for the observed data unit $(\Delta,Y,W)$. Given a marginal distribution function $F_W$ of $W$, we set $F_{P_1; W}:=F_W$ and $F_{P_2; W}:=F_W$. Given a conditional distribution function $F_{Y\midd W}$ of $Y$ given $W$, we set $F_{P_1; Y\midd W} :=F_{Y\midd W}$ and $F_{P_2; Y\midd W} :=F_{Y\midd W}$. 
Given a conditional distribution function $F_{T\midd W}$ of $T$ given $W$, we set $F_{P_1;T\midd W}=F_{T\midd W}$ and define $F_{P_2;T\midd W}$ pointwise as 
\begin{equation*}
  F_{P_2;T\midd W}(t\midd w)=m_{\alpha_2}\left[h_{\alpha_1}\left\{F_{T\midd W}(t\midd w),F_{Y\midd W}(t\midd w)\right\},F_{Y\midd W}(t\midd w)\right]\ .  
\end{equation*}
We note then that $P_1$ and $P_2$ imply, on one hand, that 
\begin{align*}
    &P_{\text{obs},1}(\Delta=1\midd Y=y, W=w)\ =\ P_1(T\leq y\midd Y=y,W=w)\\
    &\hspace{2cm}=\ h_{\alpha_1}\left\{F_{P_1;T\midd W}(y\midd w),F_{P_1;Y\midd W}(y\midd w)\right\}\\
    &\hspace{2cm}=\ h_{\alpha_1}\left\{F_{T\midd W}(y\midd w),F_{Y\midd W}(y\midd w)\right\}\ ,
\end{align*} 
and, on the other hand, that 
\begin{align*}
    &P_{\text{obs},2}(\Delta=1\midd Y=y, W=w)\ =\ P_2(T\leq y\midd Y=y,W=w)\\
    &\hspace{2cm}=\ h_{\alpha_2}\left\{F_{P_2;T\midd W}(y\midd w),F_{P_2;Y\midd W}(y\midd w)\right\}\\
    &\hspace{2cm}=\ h_{\alpha_2}\left(m_{\alpha_2}\left[h_{\alpha_1}\left\{F_{T\midd W}(y\midd w),F_{Y\midd W}(y\midd w)\right\},F_{Y\midd W}(y\midd w)\right],F_{Y\midd W}(y\midd w)\right)\\
    &\hspace{2cm}=\ h_{\alpha_1}\left\{F_{T\midd W}(y\midd w),F_{Y\midd W}(y\midd w)\right\}\ ,
\end{align*} 
implying that $P_{\text{obs},1}(\Delta=1\midd Y=y,W=w)=P_{\text{obs},2}(\Delta=1\midd Y=y,W=w)$. It can similarly be shown that $P_{\text{obs},1}(\Delta=0\midd Y=y,W=w)=P_{\text{obs},2}(\Delta=0\midd Y=y,W=w)$. Thus, the distribution of $\Delta$ given $(Y,W)$ is the same under $P_{\text{obs},1}$ and $P_{\text{obs},2}$. Of course, by definition, it is also the case that the distribution of $(Y,W)$ is the same under $P_{\text{obs},1}$ and $P_{\text{obs},2}$. Hence, we have exhibited two distributions $P_1$ and $P_2$ for the full data unit that follow the copula model with distinct indices $\alpha_1\neq \alpha_2$ but also imply the same distribution for the observed data unit. Thus, the copula parameter is not identifiable from the observed data without any additional assumption.

\section{Regression analysis}\label{sec:cox methods}
Many popular regression models used in survival analysis have been adapted for use with current status data, including the Cox proportional hazards model \citep{huang1996efficient}, the proportional odds model \citep{rossini1996semiparametric}, and the accelerated failure time model \citep{rabinowitz1995regression}. These methods are generally valid under the same assumptions on the survey response process as the adaptation of the CIR method described above. They do, however, make additional assumptions on the relationship between $T$ and $W$. 

Fitting regression models using current status data is generally more complicated than with right censored data. For the Cox model, unlike in the right-censored setting, estimation of the regression coefficients cannot be completely separated from estimation of the baseline hazard function through the partial likelihood \citep{huang1996efficient}. Nonetheless, estimation and valid inference is still attainable for the regression coefficients, with the main difficulties being computational in nature. Recent work has focused on developing more efficient algorithms for computing estimates of the regression coefficients via maximum likelihood \citep{anderson2017icenreg}. Calculation of the observed Fisher information, which is used to compute analytic standard error estimates, is generally impractical. The bootstrap has been proposed as an alternative approach \citep{pan1999extending}, although as of yet there is no theoretical proof of its validity in this context. In Section \ref{sec:sims}, we present results of a simulation study investigating the performance of the nonparametric bootstrap for construction of confidence intervals using current status data for inference on regression coefficients of a Cox model. Wald-type bootstrap confidence intervals based on a normal approximation demonstrate good performance, while percentile bootstrap confidence intervals tend to be slightly anticonservative. Based on these findings, we make use of the Wald-type interval in the regression analyses we report.

As with the Cox model, the other regression models described above do not need to be altered to account for survey nonresponse, under the assumption that the regression model is correctly specified and includes all covariates that inform the response mechanism \citep{little2019statistical}. 

\section{Simulation studies}\label{sec:sims}

\subsection{Simulations comparing extended CIR to alternative methods}\label{sec:original sims}

In this section, we describe a simulation study conducted to evaluate the performance of the proposed extended CIR accounting for survey nonresponse. While we expect the complete case CIR procedure to perform well when follow-up is unlimited, we expect decreased performance when termination of follow-up precludes survey response for some study participants. In contrast, we expect the extended CIR procedure to perform well in both of these scenarios. We also evaluated the nonparametric maximum likelihood estimator (NPMLE), which does not account for possible dependence between response times and symptom resolution, computed using only survey respondents.

In this experiment, the data were generated according to a proportional hazards model. We simulated a covariate vector $W = (W_1, W_2, W_3)$ of three independent components, each distributed according to a Unif(\{-1,1\}) distribution.  Given covariate vector $W = (w_1,w_2,w_3)$, we then simulated the survey response time $Y_*$ and event time $T$ from independent Weibull distributions with shape parameter 0.75 and scale parameters respectively set to
\begin{align*}
   &\exp(\tfrac{2}{5}w_1 - \tfrac{1}{5}w_2 + \tfrac{1}{10}w_3 + \tfrac{1}{10}w_1w_2 + \tfrac{1}{10}w_1w_3 - \tfrac{1}{10}w_2w_3)\ , \\
    &\exp(\tfrac{2}{5}w_1 - \tfrac{1}{5}w_2 + \tfrac{1}{10}w_3 + \tfrac{2}{5}w_1w_2 + \tfrac{2}{5}w_1w_3 - \tfrac{2}{5}w_2w_3)\ .
\end{align*}
For computational purposes, $Y_*$ was set to the closest point on a grid of 50 values evenly spaced on the quantile scale. In other words, the values of $Y_*$ were coarsened to 50 unique values.

For each simulation replicate, the aim was to estimate the distribution function of $T$ over the interval $[t_0, t_1]$ with $t_0 = 0.02$ (corresponding to the 5th percentile of observed response times) and $t_1 = 1.5$. The maximum follow-up time was set to $c_0 \in \{2.1, 1.8, 1.65\}$. We refer to the settings corresponding to these values of $c_0$ as Scenarios 1, 2 and 3, respectively. Under these three choices of $c_0$, $t_1$ corresponded to the 90th, 95th, or 97.5th quantile of observed survey response times, respectively. This allowed us to explore whether performance of the complete case CIR or extended CIR procedures would degrade if $c_0$ was chosen such that nonresponse was more prevalent and $t_1$ was nearer the edge of the support of observed response times. For each value of $c_0$, the observed data consisted of $n$ independent replicates of $(W, Y, \Delta)$ with $Y:=\min(Y_*, c_0)$ and $\Delta := \I(Y_* < c_0)\I(T \leq Y)$. We generated datasets of size $n \in \{500, 1000, 1500, 2000\}$. For each simulation setting, we performed 500 simulation replicates.

The survival function was estimated (1) using only complete cases (i.e., survey respondents, or those with $Y_* < c_0$) with the existing CIR procedure described in \citep{westling2020causal}, (2) using all participants with the extended CIR procedure proposed here, and (3) using the NPMLE, as implemented in the \texttt{survival} software package \citep{survival-package}, computed only using complete cases. For the two CIR procedures, nominal 95\% confidence intervals were constructed by estimating the scale factor and using the quantiles of the Chernoff distribution. The nuisance function $\mu$ was estimated using Super Learner \citep{van2007super} with a library consisting of the marginal mean, logistic regression, multivariate adaptive regression splines \citep{friedman1991multivariate}, generalized additive models (GAMs) \citep{hastie2017generalized}, and random forests \citep{wright2015ranger}. The conditional density $\pi$ was estimated using \texttt{haldensify} \citep{hejazi2022haldensify} with 10-fold cross-validation used to select optimal tuning parameters. The derivative of the survival curve was estimated by numerical differentiation of a monotone Hermite spline approximation of the estimated survival curve. For each simulation replicate, the survival function of $T$ was estimated at 100 time-points, evenly spaced on the quantile scale, between $t_0 = 0.02$ and $t_1 = 1.5$. For each setting and method, we conducted 1000 simulation replicates. Pointwise and integrated (over all 100 time-points) performance was quantified using the empirical bias and empirical confidence interval coverage. 

\begin{figure}[h!]
\centering
		\includegraphics[width=0.9\linewidth]{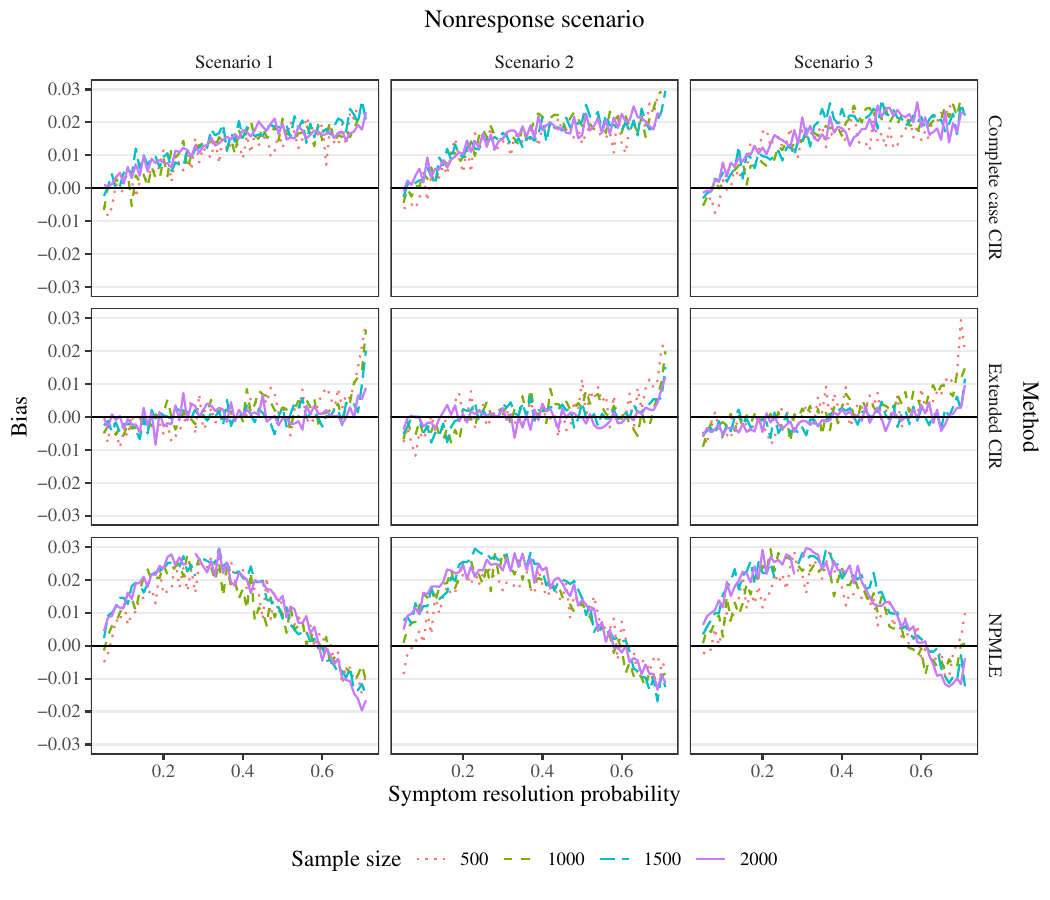}
	\caption{Pointwise bias of procedures for estimating the survival function under current status sampling with nonresponse. From left to right, the columns represent nonresponse Scenarios 1, 2 and 3, ordered from smallest to largest degree of nonresponse. From top to bottom, the rows denote the standard CIR method using only complete cases, the extended CIR method accounting for nonresponse, and the NPMLE. The $x$-axis displays the true value of the distribution function. The black line denotes zero bias.}
	\label{fig:bias plot}
\end{figure}

\begin{figure}[h!]
\centering
		\includegraphics[width=0.9\linewidth]{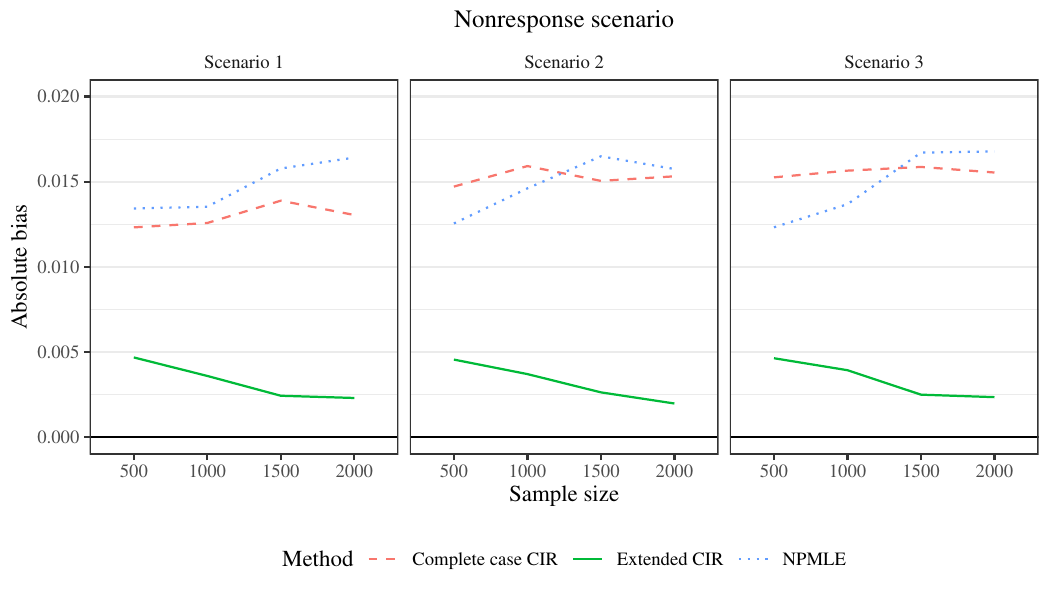}
	\caption{Integrated absolute bias of procedures for estimating the survival function under current status sampling with nonresponse. From left to right, the columns represent nonresponse Scenarios 1, 2 and 3, ordered from lowest to highest nonresponse level. The three lines in color denote the standard CIR method using only complete cases, the extended CIR method accounting for nonresponse, and the NPMLE. The $x$-axis corresponds to sample size. The black line denotes zero bias.}
	\label{fig:integrated bias plot}
\end{figure}

\begin{figure}[h!]
\centering
		\includegraphics[width=0.9\linewidth]{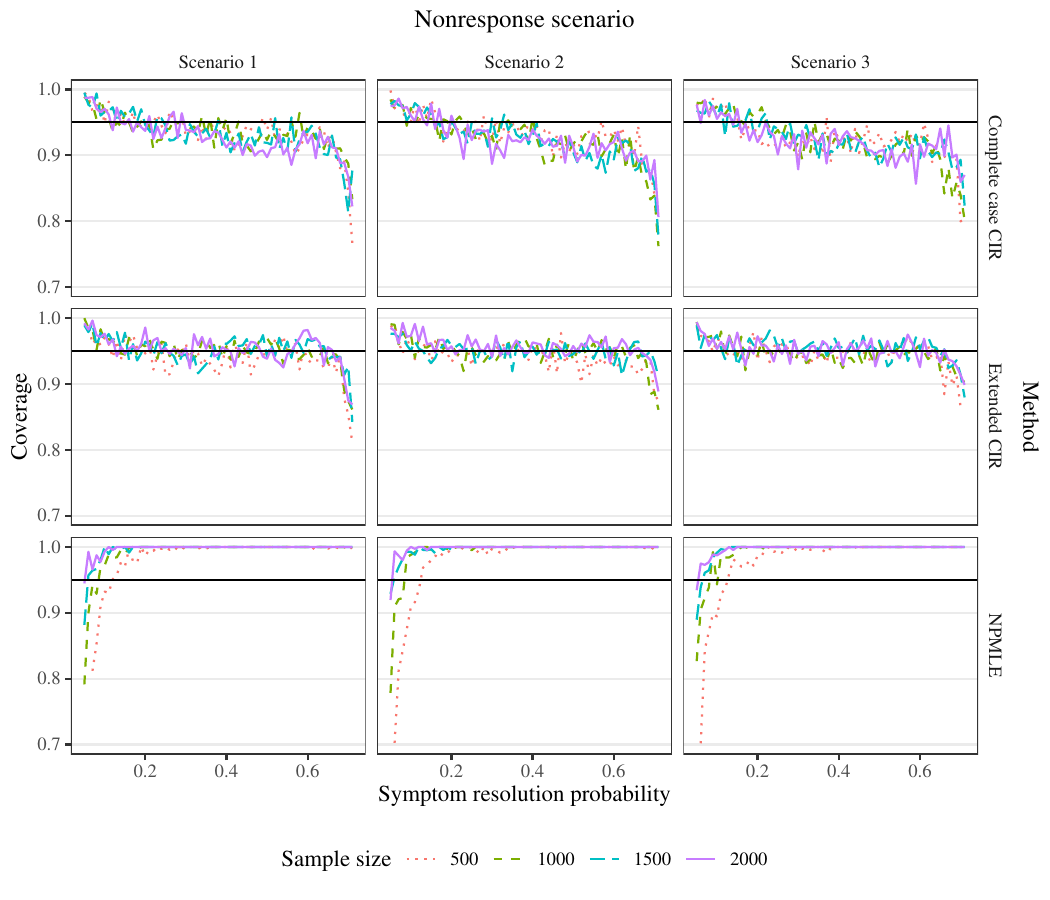}
	\caption{Confidence interval coverage using the CIR procedure for estimating the survival function under current status sampling with nonresponse. From left to right, the columns represent nonresponse Scenarios 1, 2 and 3, ordered from smallest to largest degree of nonresponse. From top to bottom, the rows denote the standard CIR method using only complete cases, the extended CIR method accounting for nonresponse, and the NPMLE. The $x$-axis displays the true value of the distribution function. The black line denotes the nominal 95\% coverage level.}
	\label{fig:coverage plot}
\end{figure}

\begin{figure}[h!]
\centering
		\includegraphics[width=0.9\linewidth]{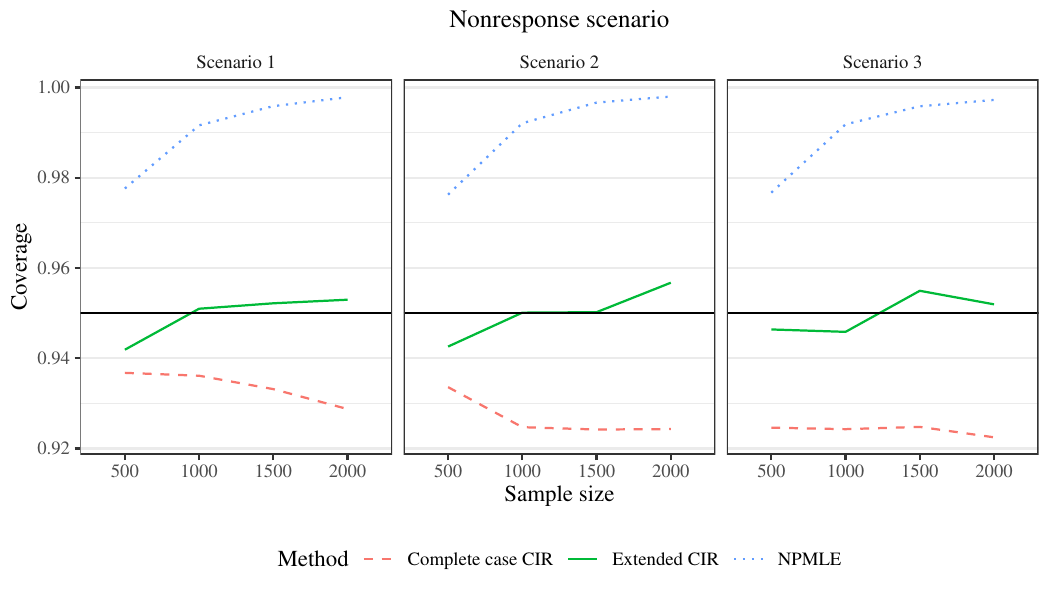}
	\caption{Confidence interval coverage using the CIR procedure for estimating the survival function under current status sampling with nonresponse. From left to right, the columns represent nonresponse Scenarios 1, 2 and 3, ordered from smallest to largest degree of nonresponse. The three lines in color denote the standard CIR method using only complete cases, the extended CIR method accounting for nonresponse, and the NPMLE. The $x$-axis corresponds to sample size. The black line denotes the nominal 95\% coverage level.}
	\label{fig:integrated coverage plot}
\end{figure}

From Figures \ref{fig:bias plot} and \ref{fig:integrated bias plot}, we see that the extended CIR generally has smaller bias compared to the complete case CIR and to the NPMLE, although all methods have generally inflated bias near the estimation boundaries. From Figures \ref{fig:coverage plot} and \ref{fig:integrated coverage plot}, we observe that the extended CIR demonstrates strong confidence interval coverage across nonresponse scenarios, while the complete case CIR tends to be somewhat anticonservative on average, and the NPMLE is highly conservative on average. 


\subsection{Robustness and stability simulations}

In this section, we describe a simulation study performed to evaluate the robustness and stability of the extended CIR procedure. Data were generated as described above, under Scenario 3, with $c_0 = 1.65$. We generated datasets of size $n \in \{500, 1000, 2500, 5000\}$.

We used the extended CIR procedure, with several different combinations of nuisance estimators, to estimate the survival function of $T$. The conditional density $\pi$ was estimated using either (1) a misspecified parametric log-normal accelerated failure time model including only linear terms, or (2) $\texttt{haldensify}$, an implementation of a nonparametric approach based on the highly adaptive lasso \citep{hejazi2022haldensify}. The outcome regression nuisance function $\mu$ was estimated using either (1) logistic regression, (2) GAMs \citep{hastie2017generalized}, (3) gradient boosted trees \citep{chen2016xgboost}, or (4) Super Learner \citep{van2007super} with a library as described in the previous experiment. For comparison, we also included the true form for $g$ and $\mu$ (i.e., in lieu of estimating the nuisance parameters, we plugged in their known values). As long as either of the two nuisance estimators is sufficiently flexible to consistently estimate its target, we expect the extended CIR procedure to yield a consistent estimator. Furthermore, if the nuisance estimation rate is fast enough, as described in Section \ref{sec:estimation and inference supplement}, we expect the resulting extended CIR estimator to not only be consistent, but to have a large-sample distribution that does not depend on the particular nuisance estimator employed. This large-sample distribution be identical as when the nuisance parameters are set to their true values. For each sample size, we conducted 500 simulations replicates; all 15 possible combinations of nuisance estimators were applied to each simulated dataset. We examined pointwise empirical bias and pointwise empirical variance. Furthermore, we examined the absolute difference between survival function estimates computed on each dataset, comparing (1) using known nuisance parameters versus using the \texttt{haldensify}--Super Learner combination, and (2) using the \texttt{haldensify}--Super Learner combination versus using the \texttt{haldensify}--gradient boosted trees combination. These two comparisons quantify the sensitivity of the proposed procedure to the choice of nuisance estimator when a flexible method is used in place of the `ideal' method using known nuisance parameter values and when two distinct flexible methods are used, respectively. 

\begin{figure}[h!]
\centering
		\includegraphics[width=0.9\linewidth]{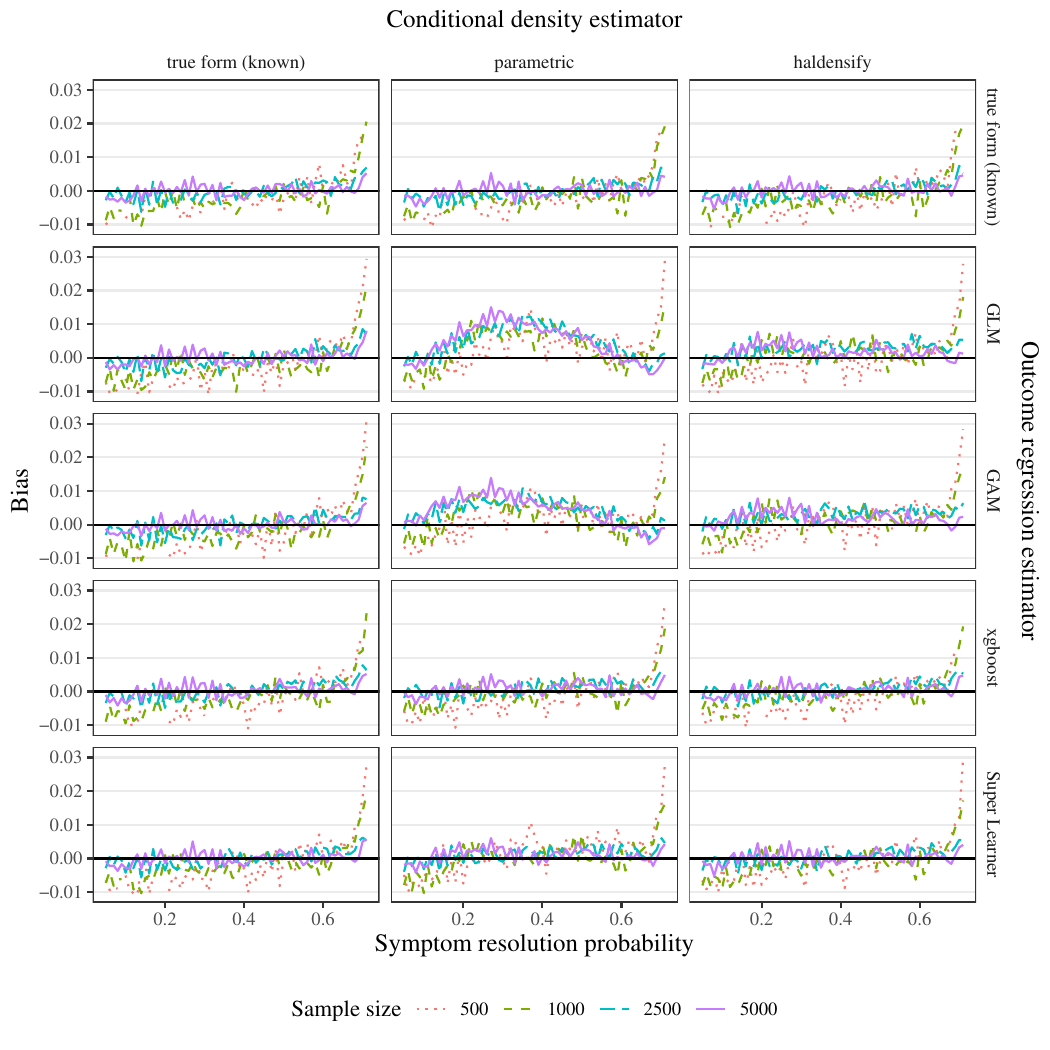}
	\caption{Pointwise bias of extended CIR procedure using various nuisance estimators. The columns represent different estimators for the conditional density $\pi$, and the rows represent different estimators for the outcome regression $\mu$. The $x$-axis displays the true value of the distribution function. The black line denotes zero bias.}
	\label{fig:stability bias plot}
\end{figure}

\begin{figure}[h!]
\centering
		\includegraphics[width=0.9\linewidth]{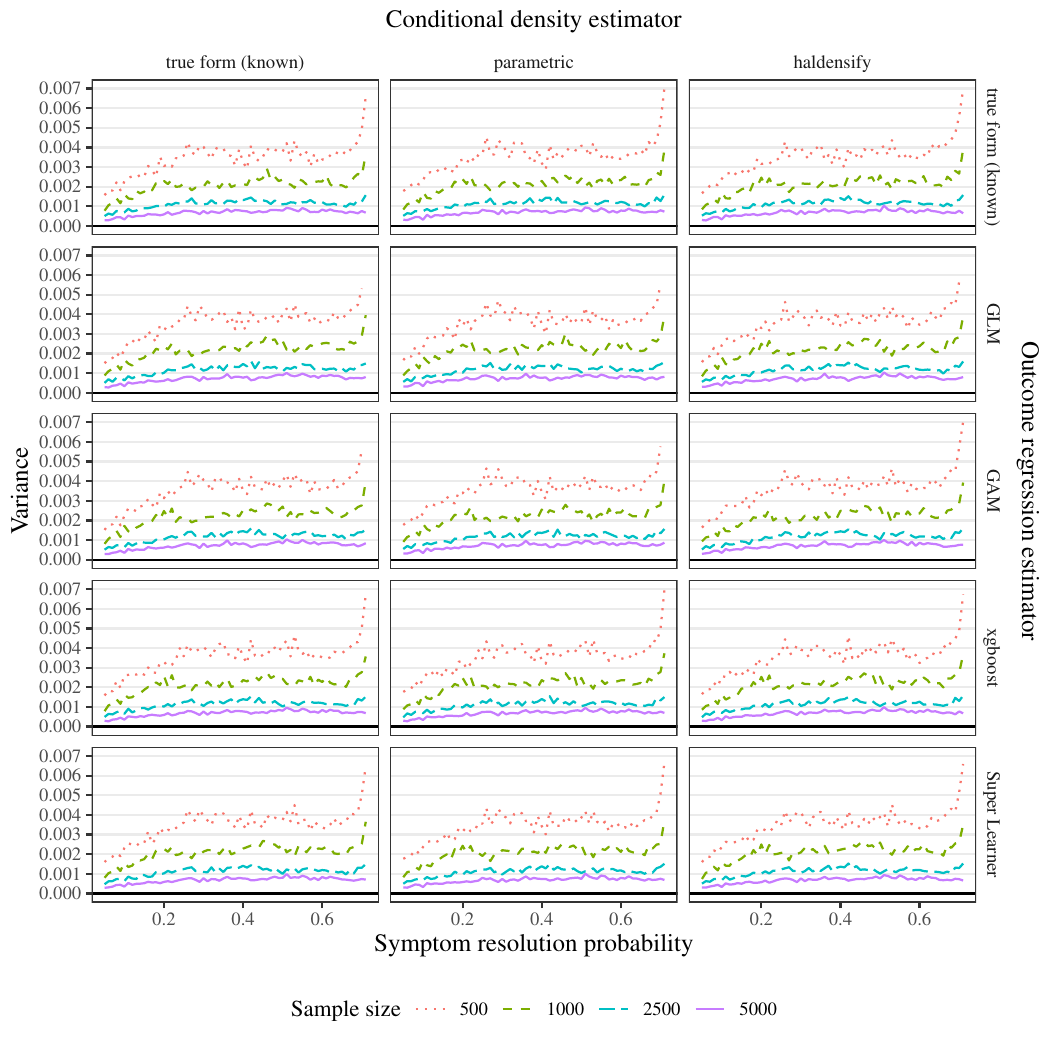}
	\caption{Pointwise variance of extended CIR procedure using various nuisance estimators. The columns represent different estimators for the conditional density $\pi$, and the rows represent different estimators for the outcome regression $\mu$. The $x$-axis displays the true value of the distribution function.}
	\label{fig:stability variance plot}
\end{figure}

From Figure \ref{fig:stability bias plot}, we observe that when $\pi$ is estimated using a misspecified parametric model and $\mu$ is estimated using either GLM or GAM, the extended CIR estimate has bias that does not shrink with increasing sample size. In these scenarios, neither of the two nuisance estimators can consistently estimate its target --- the parametric density model and GLM are both misspecified, and the underlying outcome regression is not additive --- yielding an overall CIR estimator that appears inconsistent as well. In all other cases, the overall bias is generally small. Using gradient boosted trees or Super Learner to estimate $\mu$ and \texttt{haldensify} to estimate $\pi$ results in similar performance to using the true, known values for the nuisance parameters. From Figure \ref{fig:stability variance plot}, we observe that the empirical variance of the CIR estimator exhibits similar patterns regardless of the nuisance estimators used.

\begin{figure}[h!]
\centering
		\includegraphics[width=0.9\linewidth]{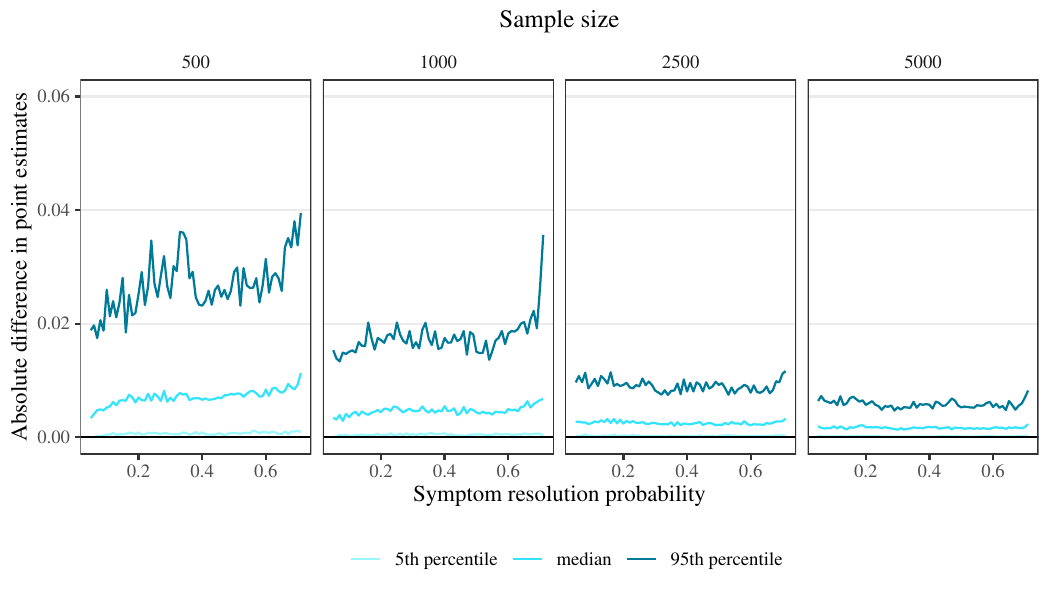}
	\caption{Pointwise absolute difference between estimates of the distribution function computed using either gradient-boosted trees or Super Learner for the outcome regression $\mu$, along with \texttt{haldensify} for the conditional density $\pi$. For each simulation replicate, each of the two nuisance estimators was applied to the simulated data and then used in the extended CIR procedure. The quantiles displayed in the plot were calculated over 500 simulation replicates.}
	\label{fig:stability plot ML}
\end{figure}

In Figure \ref{fig:stability plot ML}, we plot the median, 5th percentile, and 95th percentile absolute difference in survival function point estimates computed using \texttt{haldensify} and either gradient boosted trees or Super Learner, with the quantiles calculated across the 500 simulation replicates. This quantifies the stability of the CIR procedure; gradient boosting and Super Learner will produce different survival function estimates even when applied to the same data, but smaller differences indicate greater stability and less dependence on the chosen algorithm. We see that the stability of the procedure increases with increasing sample size. At sample size 5000, the median difference is less than 0.002 across all time points. The 95th percentile absolute difference is never larger than 0.01. Even in a relatively small sample of size 500, the median absolute difference exceeds 0.01 only near the boundary of the observable support. These results suggest that, particularly in large samples, survival estimates tend to be insensitive to the choice of nuisance estimators, at least among well-performing nuisance estimators.

\begin{figure}[h!]
\centering
		\includegraphics[width=0.9\linewidth]{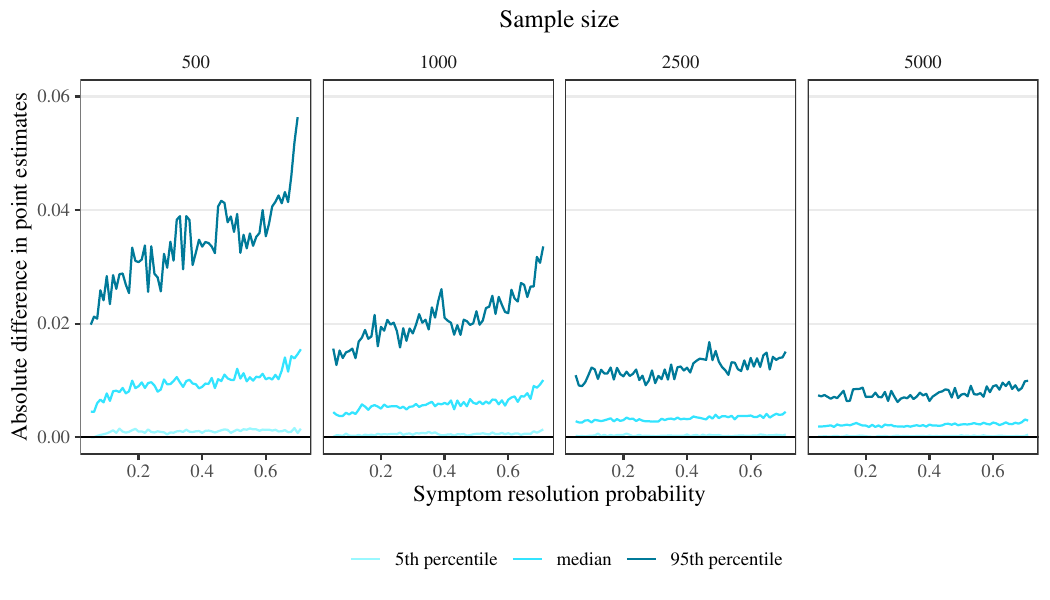}
	\caption{Pointwise absolute difference between estimates of the distribution function computed using either Super Learner and \texttt{haldensify} or the true, known forms for the outcome regression $\mu$ and conditional density $\pi$. For each simulation replicate, each of the two approaches was applied to the simulated data and then used in the extended CIR procedure. The quantiles displayed in the plot were calculated over 500 simulation replicates.}
	\label{fig:stability plot known}
\end{figure}

In Figure \ref{fig:stability plot known}, we plot the median, 5th percentile, and 95th percentile absolute difference in survival function point estimates computed using either (1) \texttt{haldensify} and Super Learner or (2) the true, known values of the nuisance parameters. Again, the stability of the procedure increases with increasing sample size, although in general the differences tend to be slightly larger than those between the two machine learning--based procedures in Figure \ref{fig:stability plot ML}. At sample size 500, the median difference does not exceed 0.02.

\subsection{Sensitivity analysis simulations}

In this section, we describe a simulation study conducted to assess the performance of the proposed sensitivity analysis approach. The data were generated according to a proportional hazards model with the Frank copula used to induce dependence between $T$ and $Y_*$ given $W$. First, we simulated a covariate vector $W$ as in the previous experiments. Given covariate vector $W = w$, we then simulated $Y_*$ from a Weibull distribution with shape parameter 0.75 and scale parameter $\exp(\tfrac{2}{5}w_1 - \tfrac{1}{5}w_2 + \tfrac{1}{10}w_3).$ Then, for association parameter $\alpha$, given $W = w$ and $Y_* = y$, the symptom resolution time $T$ was simulated according to the Frank copula with parameter $\alpha$ and Weibull distribution with scale parameter $\exp(\tfrac{2}{5}w_1 - \tfrac{1}{5}w_2 + \tfrac{1}{10}w_3)$. In this way, $T$ and $Y_*$ did not satisfy conditional independence, even given covariate vector $W$. We varied the association parameter $\alpha$ to yield Kendall's $\tau$ values
\begin{align*}
    \{-0.35,-0.25, -0.15, -0.05, +0.05, +0.15, +0.25,+0.35\}\ . 
\end{align*}
We generated datasets of size $n \in \{500, 1000, 2000, 4000\}$. The maximum follow-up time $c_0$ was set to 1.65. We aimed to estimate the cumulative distribution function of $T$ over $[t_0, t_1]$, with $t_0 = 0.02$ and $t_1 = 1.5$. 
The estimation procedure was as described in Section \ref{sec:original sims}. We compared the performance of the extended CIR procedure using the copula-based one-step estimator $\Gamma_{P_n, \alpha}(y_0) + \frac{1}{n}\sum_{i=1}^{n}D^*_{P_n,\alpha, y_0}(Z_i)$ to the performance of the extended CIR procedure using $\Gamma_{P_n}(y_0) + \frac{1}{n}\sum_{i=1}^{n}D^*_{P_n, y_0}(Z_i)$, which we expect to provide consistent estimation only under conditional independence. We refer to the latter as the `baseline' extended CIR. We quantified performance in terms of empirical bias. For each simulation setting, we performed 500 simulation replicates.

\begin{figure}[h!]
\centering
		\includegraphics[width=\linewidth]{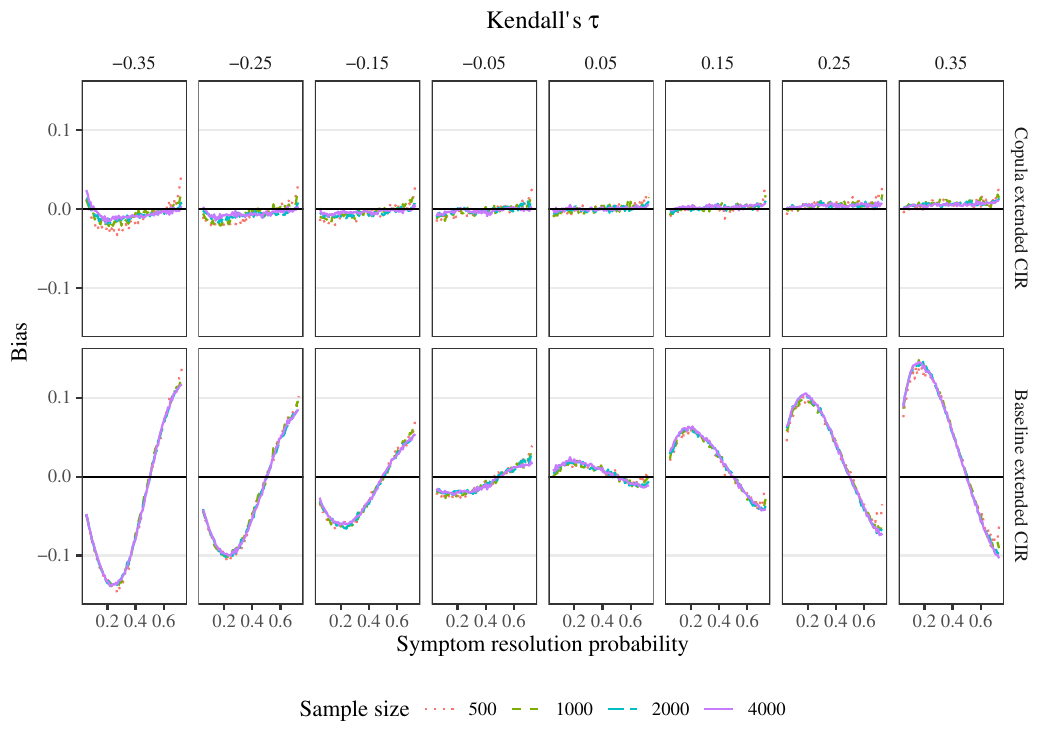}
	\caption{Pointwise bias of extended CIR procedure using the copula-based one-step estimator versus baseline extended CIR, which requires conditional independence between symptom resolution and survey response times given measured covariates. The columns represent different values of Kendall's rank correlation coefficient $\tau$, which describes the level of dependence between symptom resolution time and survey response time, conditional on covariates, with $\tau = 0$ representing conditional independence. The $x$-axis displays the true value of the distribution function. The black line denotes zero bias.}
	\label{fig:sensitivity sim}
\end{figure}

From Figure \ref{fig:sensitivity sim}, we observe that the copula-based estimator has substantially reduced bias compared to the baseline extended CIR estimator. The magnitude of excess bias for the baseline extended CIR estimator increases for values of Kendall's $\tau$ further from 0.

\subsection{Cox proportional hazards regressions simulations}
In this section, we describe a simulation study performed to evaluate the validity of the nonparametric bootstrap for constructing confidence intervals for regression coefficients of the Cox proportional hazards model when the data are subject to current status sampling and survey nonresponse. While this is not meant to replace rigorous theoretical analysis, it may help inform the practical use of the bootstrap in this setting. 

\begin{figure}[h!]
\centering
		\includegraphics[width=0.9\linewidth]{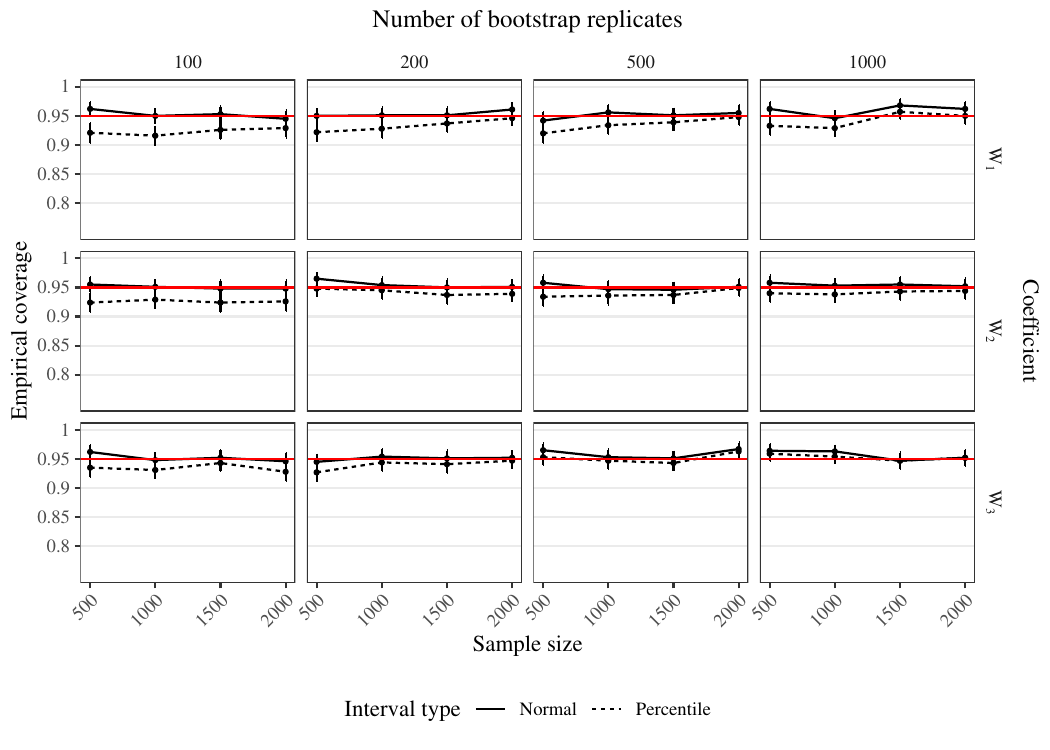}
	\caption{Performance of nonparametric bootstrap confidence intervals for the Cox model coefficients under current status sampling and nonresponse in Scenario 1. The columns represent different numbers of bootstrap replicates, and the rows represent the regression coefficients for the three covariates included in the model. In each plot, the solid black line shows the confidence interval coverage of nominal 95\% Wald-type confidence intervals based on a Normal approximation, while the dashed black line shows the coverage of nominal 95\% percentile confidence intervals. The red line denotes 95\% coverage.}
	\label{fig:bootstrap scen1}
\end{figure}

\begin{figure}[h!]
\centering
		\includegraphics[width=0.9\linewidth]{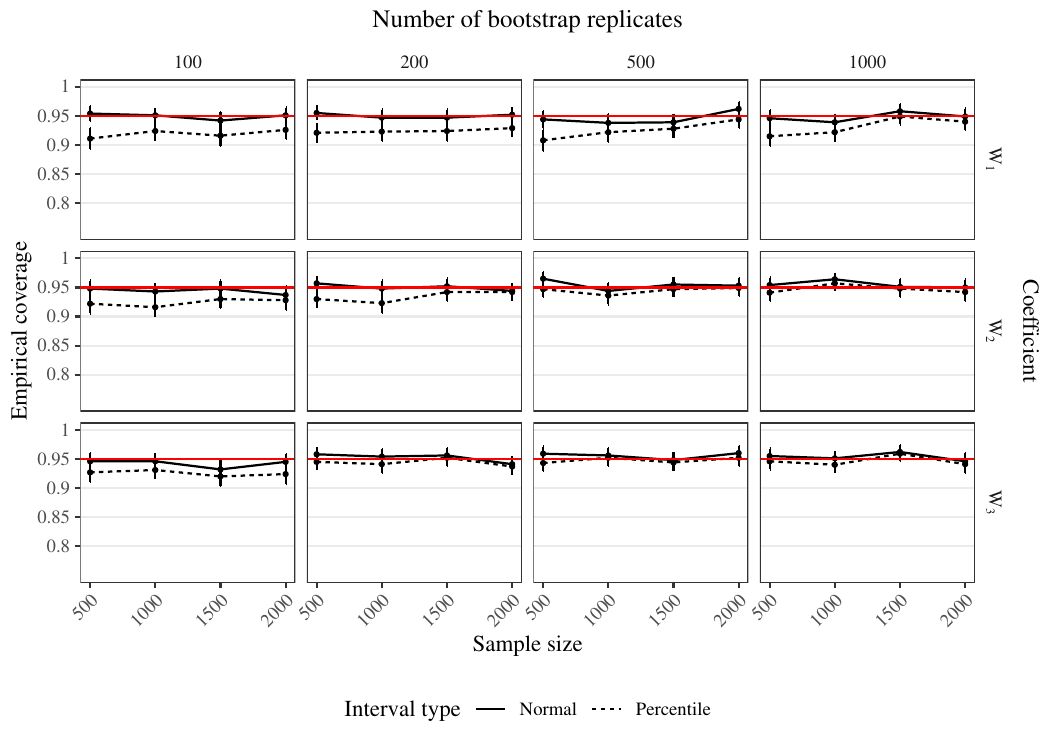}
	\caption{Performance of nonparametric bootstrap confidence intervals for the Cox model coefficients under current status sampling and nonresponse in Scenario 2. The columns represent different numbers of bootstrap replicates, and the rows represent the regression coefficients for the three covariates included in the model. In each plot, the solid black line shows the confidence interval coverage of nominal 95\% Wald-type confidence intervals based on a Normal approximation, while the dashed black line shows the coverage of nominal 95\% percentile confidence intervals. The red line denotes 95\% coverage.}
	\label{fig:bootstrap scen2}
\end{figure}

\begin{figure}[h!]
\centering
		\includegraphics[width=0.9\linewidth]{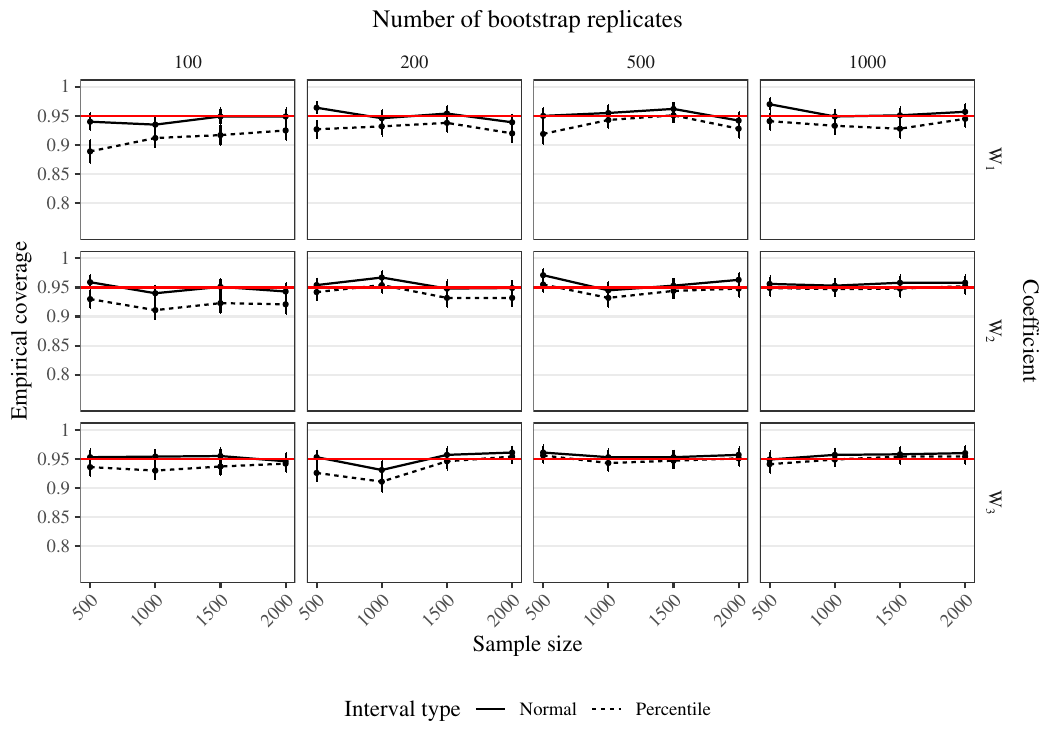}
	\caption{Performance of nonparametric bootstrap confidence intervals for the Cox model coefficients under current status sampling and nonresponse in Scenario 3. The columns represent different numbers of bootstrap replicates, and the rows represent the regression coefficients for the three covariates included in the model. In each plot, the solid black line shows the confidence interval coverage of nominal 95\% Wald-type confidence intervals based on a Normal approximation, while the dashed black line shows the coverage of nominal 95\% percentile confidence intervals. The red line denotes 95\% coverage.}
	\label{fig:bootstrap scen3}
\end{figure}

Data were generated similarly as in the previous set of experiments; however, the scale parameter of the Weibull distribution was set to $\exp(\tfrac{2}{5}w_1 - \tfrac{1}{5}w_2)$, satisfying a linear proportional hazards model. To evaluate the performance of the bootstrap, we fit a correctly specified Cox model (i.e., an additive model with only first-order terms) on each of $B$ bootstrap resamples of the dataset. We constructed confidence intervals for each regression coefficient using either a normal approximation with standard deviation equal to the standard deviation of the point estimates across the $B$ replicates or a percentile bootstrap. We repeated this procedure 1000 times and calculated the empirical confidence interval coverage for each method of interval construction. We generated datasets of size $n\in\{500,1000,1500,2000\}$ and used bootstrap replicates $B\in\{100,250,500,1000\}$. For each simulation setting, we performed 500 simulation replicates. The empirical coverage for the coefficients corresponding to $W_1,W_2$  and $W_3$ in each of these settings is displayed in Figures \ref{fig:bootstrap scen1}--\ref{fig:bootstrap scen3}. 
	
From Figures \ref{fig:bootstrap scen1}--\ref{fig:bootstrap scen3}, we observe that, under all nonresponse mechanisms, the normal approximation bootstrap performs well, with empirical coverage within Monte Carlo error of the nominal level of 0.95 across both sample sizes and number of bootstrap replicates. The percentile bootstrap produces somewhat anticonservative confidence intervals, although its performance tends to improve with a larger number of bootstrap replicates.


\section{Supplementary information for HCT data analysis}\label{sec:data analysis supp}
This section contains supplementary information on the analysis of the HCT long COVID survey data.

\subsection{Eligibility criteria for HCT enrollment}
The eligibility criteria were 
\begin{enumerate}
    \item a valid university identification number;
    \item age 13 years or older; 
    \item working or attending classes in-person at the university campus at least once per month
    \item living within commuting distance of one of the three campuses
    \item ability to provide consent in English.
\end{enumerate}
Any individual who met the inclusion criteria could enroll in the parent HCT study. 

\subsection{Algorithm library for Super Learner}

For data analyses and simulations Super Learner \citep{van2007super} was implemented with a library consisting of the marginal mean, logistic regression, multivariate adaptive regression splines \citep{friedman1991multivariate}, generalized additive models \citep{hastie2017generalized}, random forests \citep{wright2015ranger}, and gradient boosted trees \citep{chen2016xgboost}. 

\subsection{Supplementary figures}

\begin{figure}[h!]
\centering
		\includegraphics[width=\linewidth]{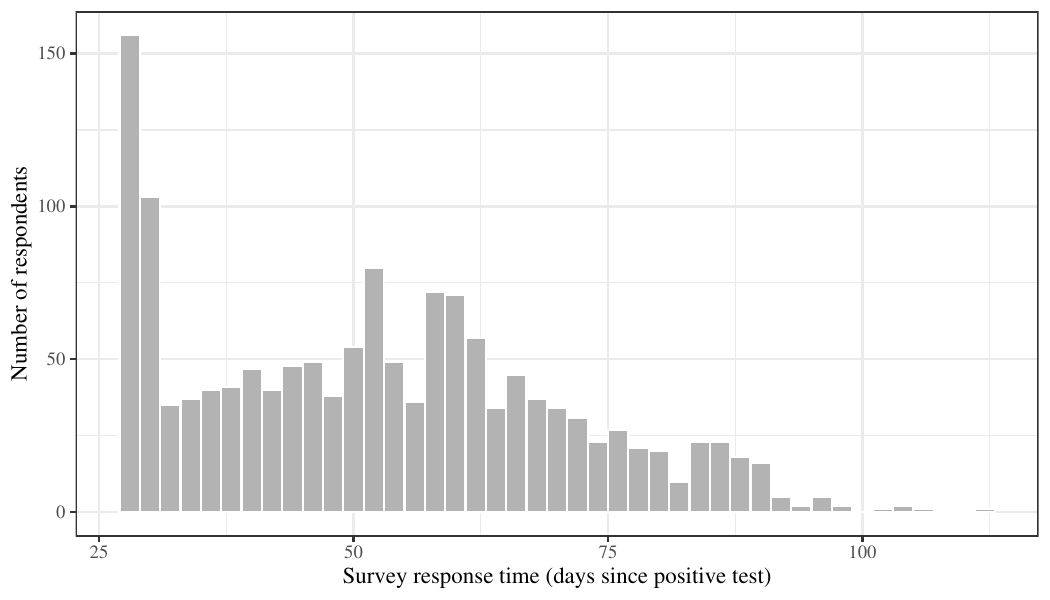}
	\caption{Histogram of survey response times. Surveys were sent at least 28 days after the positive test.}
	\label{fig:response times histogram}
\end{figure}

\begin{figure}[h!]
\centering
		\includegraphics[width=\linewidth]{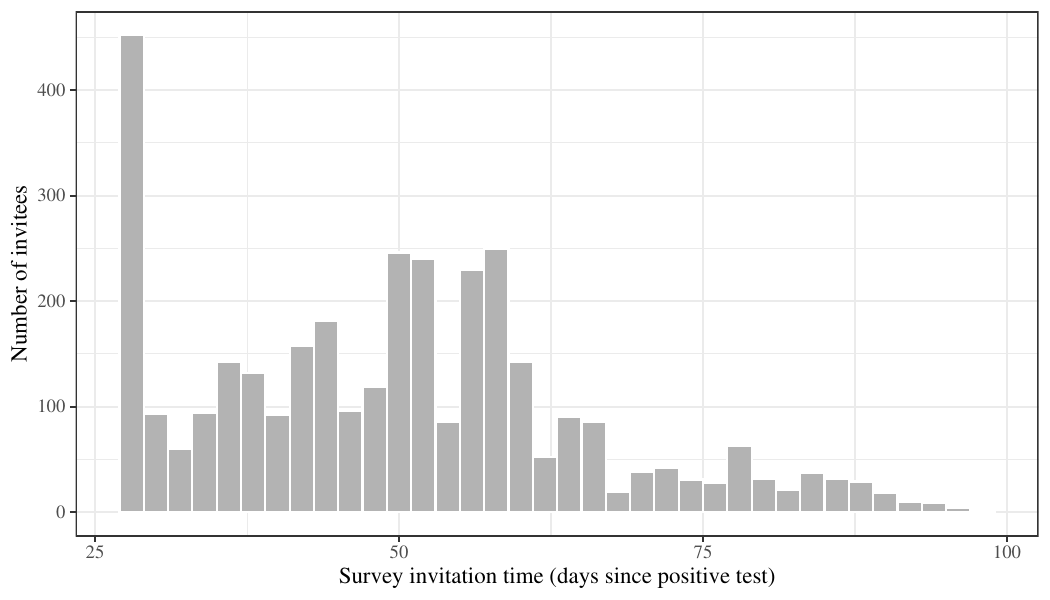}
	\caption{Histogram of survey invitation times smaller than 120 days. Surveys were sent at least 28 days after the positive test, and 33 invitation times were larger than 120 days.}
	\label{fig:invite lag times histogram}
\end{figure}

Figure \ref{fig:response times histogram} shows the distribution of survey response time $Y$ among HCT survey respondents. Figure \ref{fig:invite lag times histogram} shows the distribution of survey invitation time (time between positive test and survey invitation being sent) among all HCT survey invitees.

\end{document}